\documentclass[review, 10pt]{elsarticle}

\journal{Journal of Computational Statistics \& Data Analysis}

\usepackage{hyperref}
\usepackage{times}
\usepackage{epsfig}
\usepackage{graphicx}
\usepackage{amsmath}
\usepackage{amssymb}
\usepackage{booktabs}
\usepackage{algorithm}
\usepackage{algpseudocode}
\usepackage[T1]{fontenc}
\usepackage[utf8]{inputenc}
\usepackage{hyperref}
\usepackage{bigstrut}
\usepackage{subcaption}
\usepackage{multirow}
\usepackage{multicol}
\usepackage{changepage}
\usepackage{xcolor}
\usepackage{caption}

\def\argmin{\mathop{\rm argmin}}

\hypersetup{breaklinks=true,
            bookmarks=true,
            colorlinks=true,
            citecolor=blue,
            urlcolor=blue,
            linkcolor=magenta,
            pdfborder={0 0 0}}

\newcommand{\inner}[2]{\left\langle #1,#2 \right\rangle}

\newcommand{\real}{\ensuremath{\mathbb{R}}}

\newcommand{\ltwo}{\ensuremath{\mathbb{L}^2}}

\newcommand{\ie}{{\it i.e.}}

\begin{document}

\begin{frontmatter}

\title{Regression Models Using Shapes of Functions as Predictors}

\author{Kyungmin Ahn}
\address{RIKEN Center for Biosystems Dynamics Research (BDR), Kobe, Japan}
\author{J. Derek Tucker}

\address{Sandia National Laboratories, Albuquerque, New Mexico, USA}

\author{Wei Wu, Anuj Srivastava}
\address{Florida State University, Tallahassee, Florida, USA}

\begin{abstract}
Functional variables are often used as predictors in regression problems. A
commonly used parametric approach, called {\it scalar-on-function regression}, uses the $\ltwo$ inner product to 
map functional predictors into scalar responses. This method can perform poorly when predictor functions contain 
undesired phase variability, causing phases to have disproportionately large influence on the response variable. 
One past solution has been to perform phase-amplitude separation (as a pre-processing step) and then 
use only the amplitudes in the regression model. Here we propose a more integrated approach, termed {\it elastic functional regression model} (EFRM), 
where phase-separation is performed inside the regression model, rather than as a pre-processing step. 
This approach generalizes the notion of phase in functional data, and is based on 
the norm-preserving time warping of predictors. Due to its invariance properties, this
representation provides robustness
to predictor phase variability and results in improved predictions of the response variable over traditional 
models. We demonstrate this framework using a number of datasets involving gait signals, 
NMR data, and stock market prices.
\end{abstract}


\end{frontmatter}

\section{Introduction}

A fast growing subtopic in functional data analysis (FDA) \cite{ramsay-silverman-2005} is 
{\it regression} involving functional 
variables, either as predictors or responses or both.
Morris \cite{morris:2015} categorizes regression problems involving functional data into three types: (1) functional predictor regression (scalar-on-function), (2) functional response regression (function-on-scalar) and (3) function-on-function regression. 
The functional predictor regression problem (or scalar-on-function) model was first studied by Ramsay \cite{ramsay-dalzell:1991}, 
Cardot et al. \cite{cardot-ferraty-sarda:1999}, and several other 
since then \cite{ahn2018elastic, james:2002, reiss2017, goldsmith2014estimator, fuchs2015penalized, ciarleglio2016wavelet, gertheiss2013longitudinal,cai2006prediction}. 
In this set up, predictors are 
scalar-valued functions over a fixed interval say $[0,T]$,
call them $\{f_i \in {\cal F} \}$, elements of some pre-specified functional space ${\cal F}$, and responses 
are scalars $\{y_i \in \real\}$ (One can easily extend this framework to 
the case where predictors or responses are vector-valued). 
A simple and commonly-used model for this problem is the so-called 
{\it functional linear regression model} (FLM) given by:
\begin{equation}
	y_i = \alpha + \inner{\beta}{f_i}+\epsilon_i,~~i=1,\dots,n\ ,
	\label{eq:regress_model}
\end{equation}
where $\alpha \in \real$ is the intercept, $\beta \in {\cal F}$ is the regression-coefficient function, and 
$\epsilon_i \in \real$ is the observation noise. Also, 
$\inner{\beta}{f_i}$ denotes the standard $\ltwo$ inner product $ \int_0^T f_i(t)\beta(t)\,dt$. 
(Notationally, we will use $\| \cdot \|$ to denote the $\ltwo$ norm.) 
One assumes here that ${\cal F}$ has the 
$\ltwo$ Hilbert structure to allow for this 
inner product between its elements. 
Similar to linear regression models with Euclidean variables, one can also 
estimate model parameters here by minimizing the sum of squared errors 
(SSE):
\begin{equation}
\{\hat{\alpha}, \hat{\beta}\} = \argmin_{\alpha \in \real,\beta \in \ltwo} \left[ \sum_{i=1}^n \left(y_i - \alpha - \inner{\beta}{f_i} \right)^2 \right]\ .
\label{eq:ols}
\end{equation}
However, since $\ltwo$ is infinite dimensional, this problem is not sufficiently constrained 
to estimate $\hat{\beta}$ with a 
finite sample size $n$, and requires further
restrictions. These constraints can come in form of a regularization term 
or a restriction of the solution space, or both.
For restricting the solution space, one can use a complete orthonormal basis of ${\cal F}$, for representing $\beta$ 
via its coefficients, and then 
truncate it to make the representation finite dimensional. A regularization is often imposed using
a roughness measure on $\beta$, e.g. $\int \ddot{\beta}(t)^2 dt$ (For a function $f(t)$, we will 
use $\dot{f}(t)$ and $\ddot{f}(t)$ to denote its first and the second derivatives, respectively).
The FLM model can easily be extended to a {\it generalized FLM} 
\cite{muller2005generalized}, where the conditional mean of the 
response given the predictors uses a known link function.
\\

\subsection{Basic Issue: Predictor Phase}
While the use of functional data has grown in recent years, there has also been a growing awareness 
of a problem/issue that is specific to functions. 
Functional data often comes with a {\it phase variability}, {\it i.e.} a lack 
of registration between geometric features (peaks, valleys, etc.)  
across functions \cite{marron2015functional,marron2014statistics,srivastava-etal-function:2011}.
Different observations can 
potentially represent different 
temporal rates of evolutions, introducing an intrinsic phase variability in the data. 
This situation arises, for example, in biological signals, growth curves, pandemic curves, 
and stock market data. In all these examples,
functional measurements often lack temporal synchronizations across measurements.

In mathematical terms, the functional data is not $\{ f_i\}$, as in the original model,  
but observed under random time warpings. Let  $\Gamma$ be the set of all time warping functions (formally 
defined later). 
In fact, depending on the context, three types of warpings are possible.
\begin{itemize}
\item {\bf Value-preserving warping}:  The most commonly-used mapping is
$f_i \mapsto (f_i \circ \gamma_i)$. It is called {\it value-preserving warping} 
as it preserves the heights of the function $f_i$ and only 
shifts them horizontally. It is often used in the alignment of peaks and valleys in functional data. 

\item {\bf Area-preserving warping}:
The mapping 
 $f_i \mapsto (f_i \circ \gamma_i){\dot{\gamma}_i}$, is called an {\it area-preserving warping}, 
 since it preserves the area under the curve $f_i$. It is often used when $\{f_i\}$ are probability 
 density functions. 

\item {\bf Norm-preserving warping}:
Another warping results from the mapping 
 $f_i \mapsto (f_i \circ \gamma_i) \sqrt{\dot{\gamma}_i}$, called a {\it norm-preserving warping}, 
 since it preserves the $\ltwo$ norm of $f_i$. That is, $\|f_i\| = \| (f_i \circ \gamma_i) \sqrt{\dot{\gamma}_i}\|$ for 
 all $f_i \in \ltwo$ and $\gamma \in \Gamma$. 
\end{itemize}
Additional types of warpings may also be possible, depending on the 
need of the application. While in functional data alignment, one mainly uses the value-preserving warping of functions, 
we will keep our options more general in this paper. Since our goal is regression and prediction, not
just functional alignment, we are free to incorporate any type of warping in the model, as needed. 
In the following, we will use $(f_i^* \gamma_i)$ as an encompassing notation for  
above-mentioned warpings.

\begin{figure}[htb!]
\begin{center}
\begin{tabular}{|c|cc|}
\hline
\includegraphics[height=1.1in]{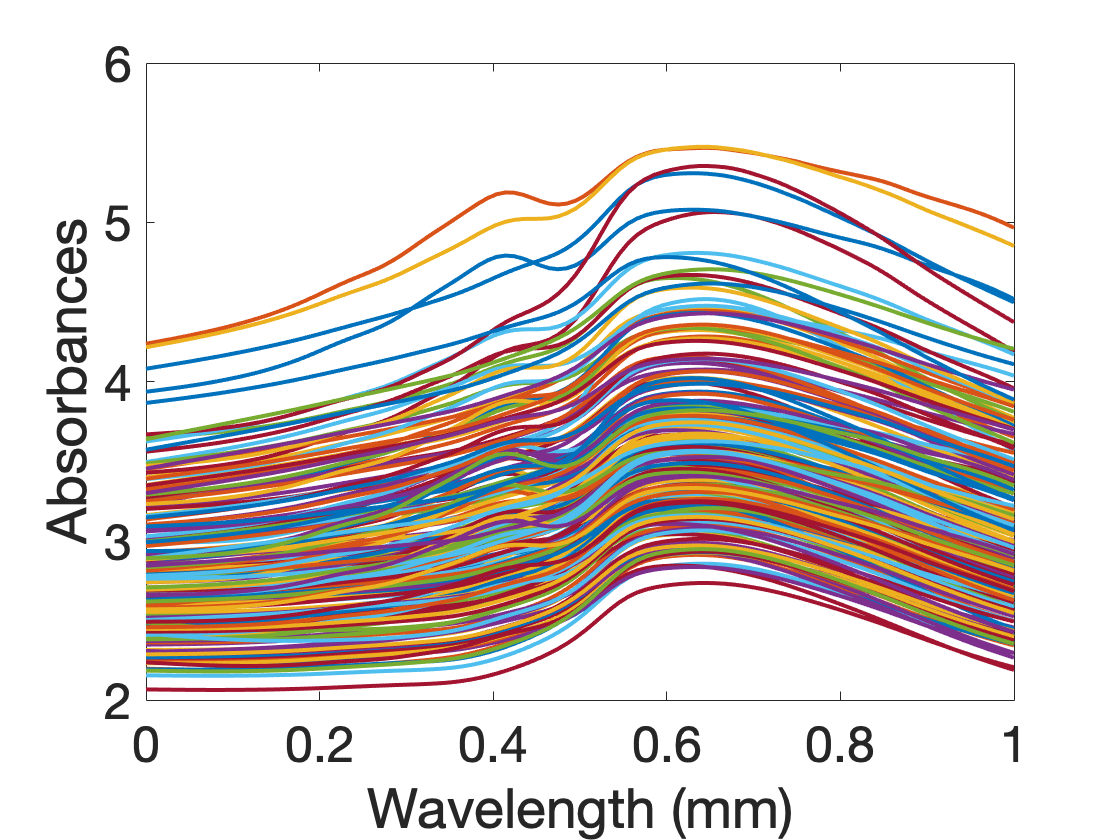} &
\includegraphics[height=1.1in]{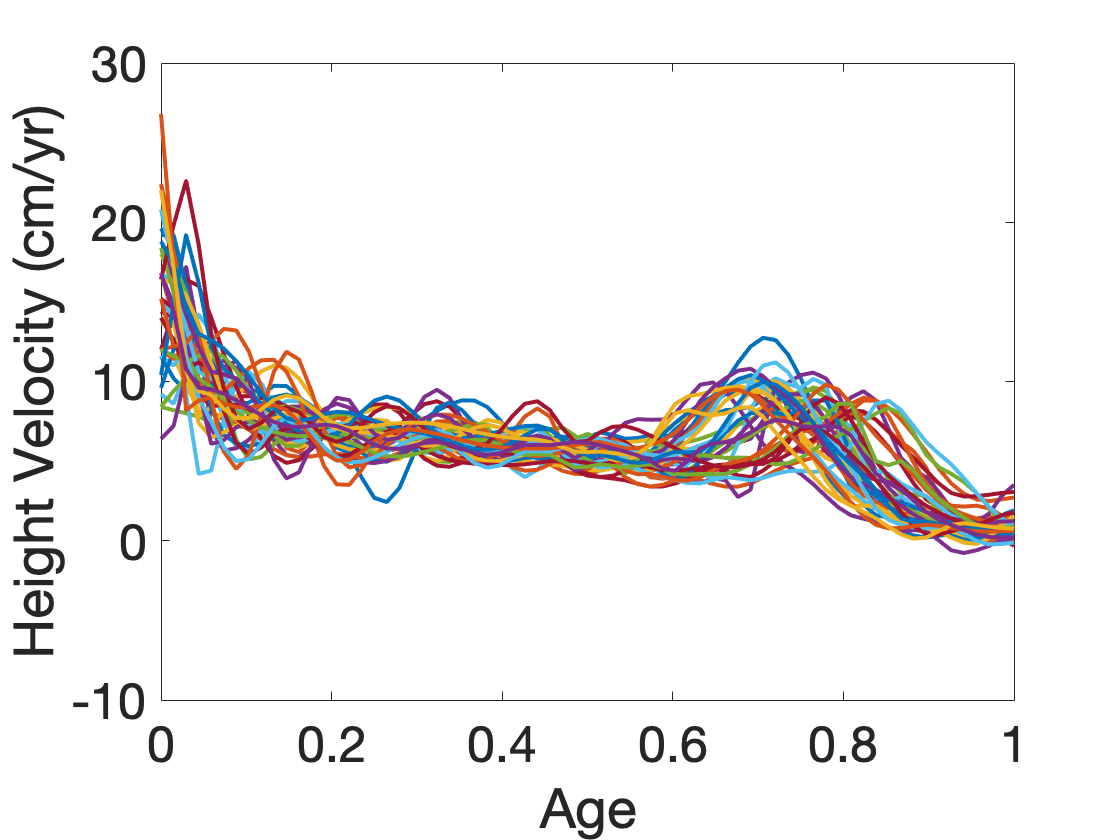} &
\includegraphics[height=1.1in]{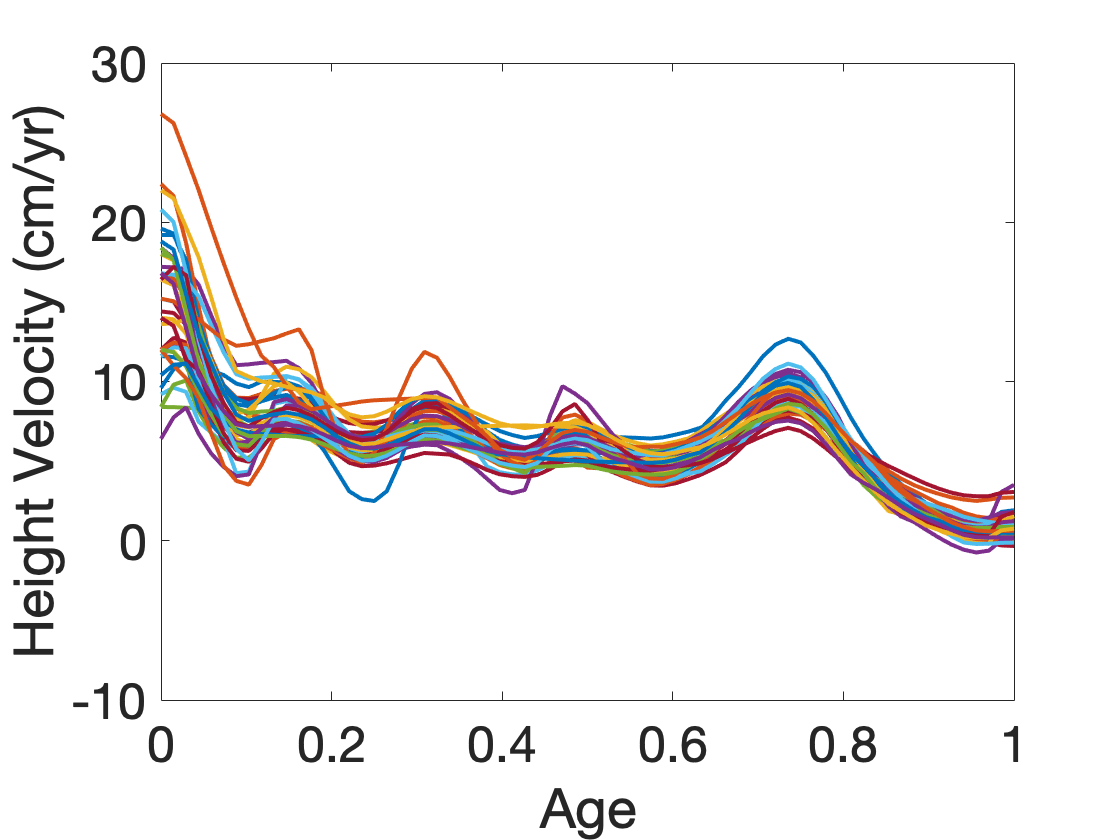} \\
 & Original & Aligned \\
Tecator Benchmark Data & \multicolumn{2}{c|}{Berkeley Growth Data} \\
\hline
\end{tabular}
\caption{Example of functional data with and without phase variability.}
\label{ex1}
\end{center}
\end{figure}

In FDA, it is often advantageous and sometimes imperative
to take into account time warpings of functional data. 
Examples of such treatments in data analysis  include 
\cite{marron2015functional,marron2014statistics,srivastava-etal-function:2011} and 
in data modeling include \cite{tucker-etal:2013}. 
For instance, a common idea in FDA is to perform
alignment of peaks and valleys across functions using the value-preserving warpings ($f_i^*\gamma_i = f_i \circ \gamma_i)$
of their domains. These warpings $\{ \gamma_i\}$ correspond 
to the {\it phase} components and the aligned functions $\{ f_i \circ \gamma_i\}$ correspond to the 
{\it shape} or the {\it amplitude} components. 
To illustrate this, consider two data examples shown in Fig. \ref{ex1}. 
On the left, we see the {\it Tecator} data which shows 
absorbance curves for certain meat and has been used 
commonly in several FDA papers \cite{febrero2012statistical, garcia2014goodness}. The predictor 
functions here are already well registered and one can use them directly in a 
statistical model without any consideration of phase. 
The right
side shows a different situation, involving the famous {\it Berkeley growth} data, where height changes 
of 69 male subjects are 
displayed in the middle panel. While these curves have a similar number of peaks and valleys, these features are 
not well aligned across subjects, due to differences in growth rates and the body clocks across subjects. 
Since this data contains a larger phase variability, 
the problem of phase-amplitude separation becomes important. 
The result of one such alignment algorithm \cite{srivastava-etal-function:2011} applied to the data is
shown in the right panel. As the reader can see, the peaks and valleys in functions are now well aligned. 

In general regression models both components of predictors -- phase and shape -- are useful.
However, there are situations where only one of them, most notably, the shape, 
is of interest in predicting a response variable. This situation arises, for instance, in cases where the 
response depends primarily on the numbers and heights of the modes in  
the predictor functions, and the {\it locations} of modes and anti-modes are not influential and are 
considered nuisance. 
To motivate this further, using the human growth data, imagine a certain 
response variable, say the gender of the subject, that depends primarily on shapes of these curves
and not on the locations of growth spurts. Thus, shape-based functional regression 
becomes a useful tool in this context.
Motivated by such problems, we shall develop a regression model where only the shape (or amplitude)
of a function is used in the model and its phase is removed from the consideration. 

The phase variability in functional predictors, even if small, can have a disproportionately large influence on statistical analysis. 
One consequence of 
phase variability is the inflation of variance in the predictor itself, {\it i.e.}
the variance of $\{ (f_i^*\gamma_i) \}$ can be much higher than that of $\{f_i\}$, 
rendering any ensuing variance-based analysis ineffective. 
Another consequence is the change in the regression model itself. 
Under the value-preserving warping, using the Taylors' expansion, we get
$$
f_i(\gamma_i(t)) = f_i(\gamma_{id}(t)) + \dot{f}_i(t) (\gamma_i(t) - \gamma_{id}(t)) + \ \ \mbox{higher order terms}\ ,
$$
with $\gamma_{id}(t) = t$.
Dropping the higher-order terms and replacing $f_i$ by $f_i \circ \gamma_i$ in Eqn.~\ref{eq:regress_model}, we get 
$$
E[y_i| \beta, f_i] = \alpha + \inner{\beta}{f_i} + \inner{\beta}{\dot{f}_i \cdot (\gamma_i - \gamma_{id})}\ .
$$
The conditional mean gets changed, up to the first order, by an amount captured by the third term on the right side. 
Depending on the value of $\{\dot{f}_i\}$, this change can be significant, adversely affecting the prediction 
performance. Although this derivation involves value-preserving time warping, a similar analysis can 
be repeated for other group actions also, with similar conclusions. 
Sometimes the phase variability are simple linear or affine shifts, and can be handled 
trivially, 
but in general phases are nonlinear functions and require more comprehensive 
mathematical tools. 

\begin{figure}
\begin{center}
\begin{tabular}{cccc}
\includegraphics[height=1.1in]{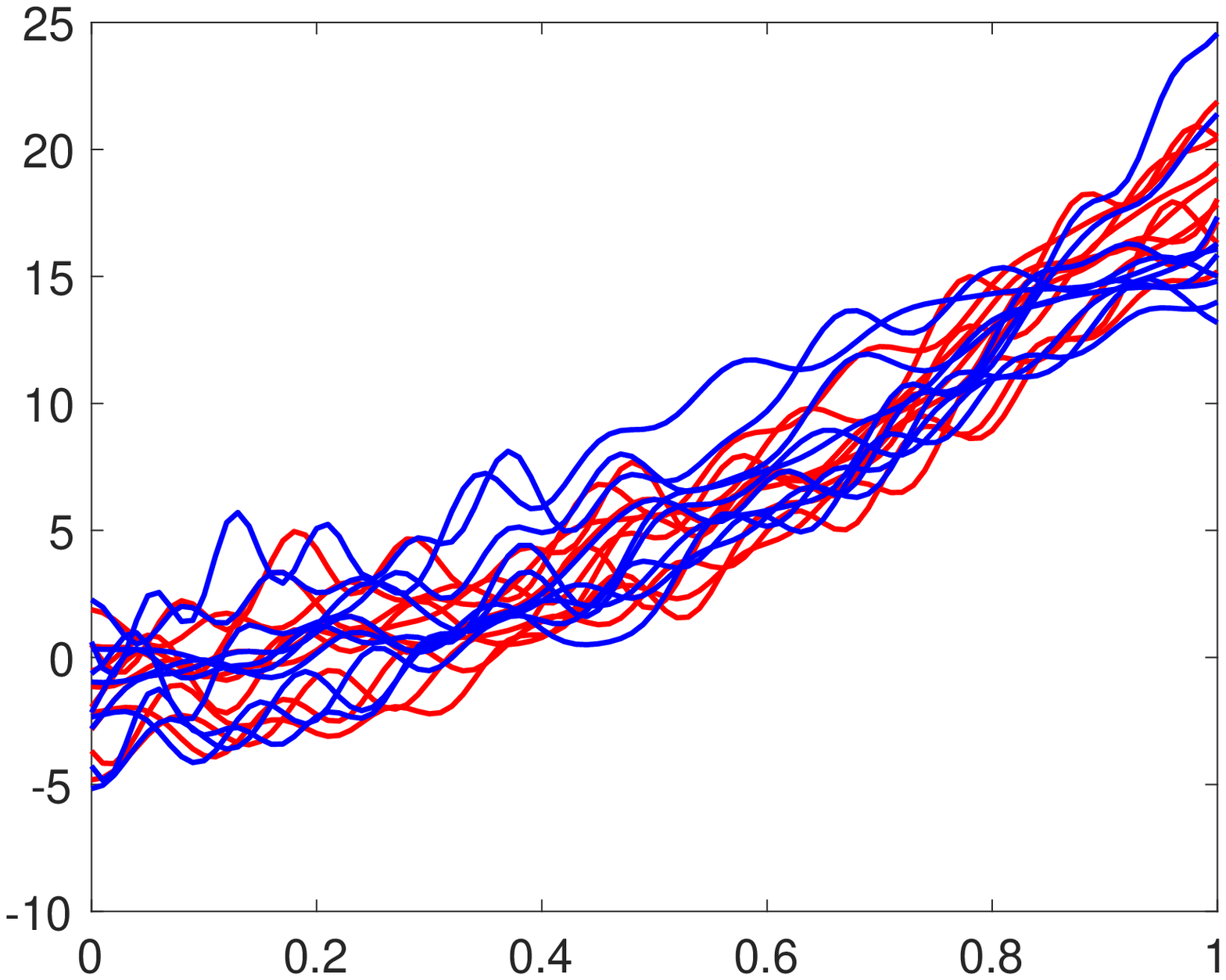} &
\includegraphics[height=1.1in]{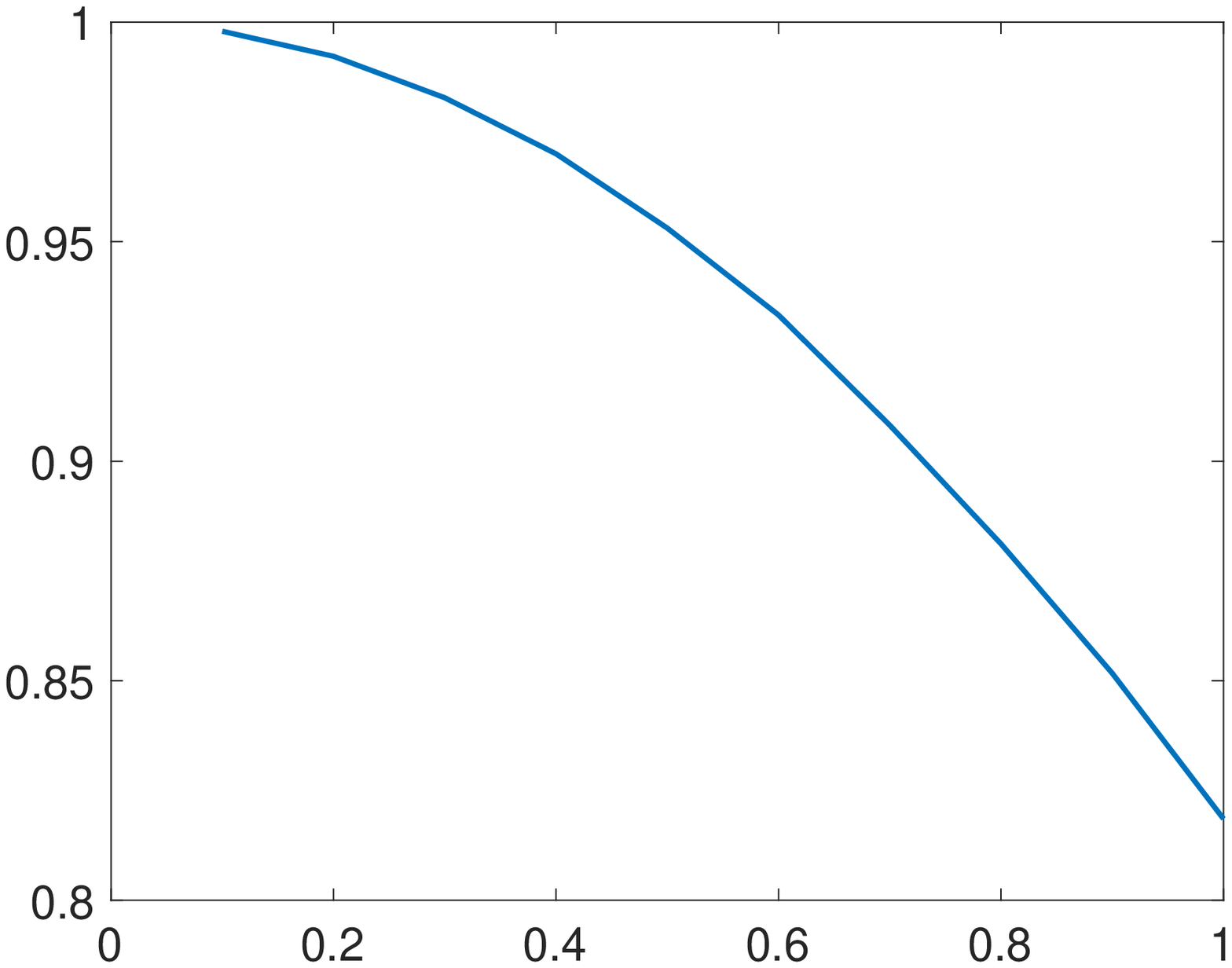} &
\includegraphics[height=1.1in]{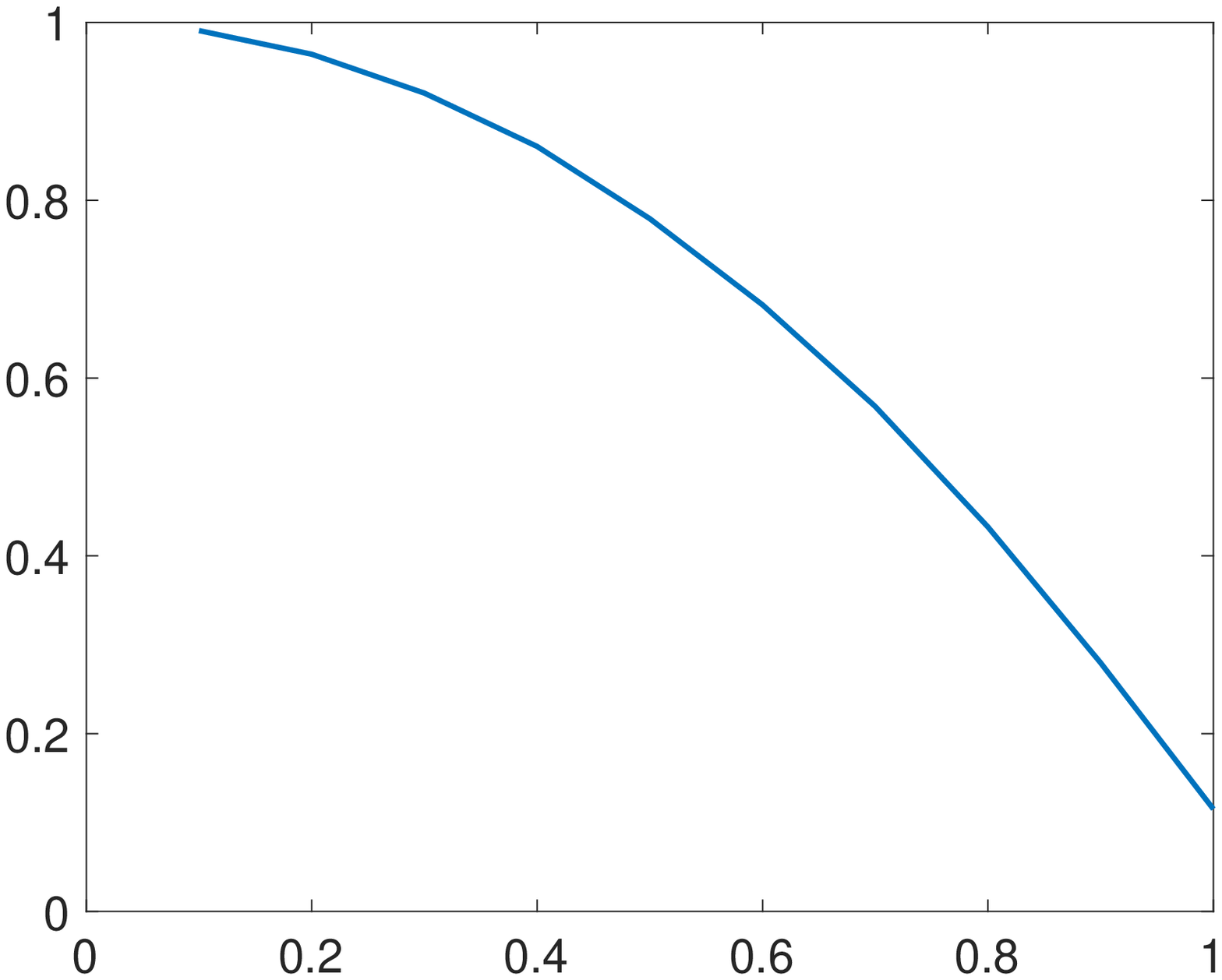}\\
$\{f_i\}$ & $R^2$ versus $a$, $(f_i \circ \gamma_i) \sqrt{\dot{\gamma}_i}$ &  $R^2$ versus $a$, ($f_i \circ \gamma_i$)\\
\includegraphics[height=1.1in]{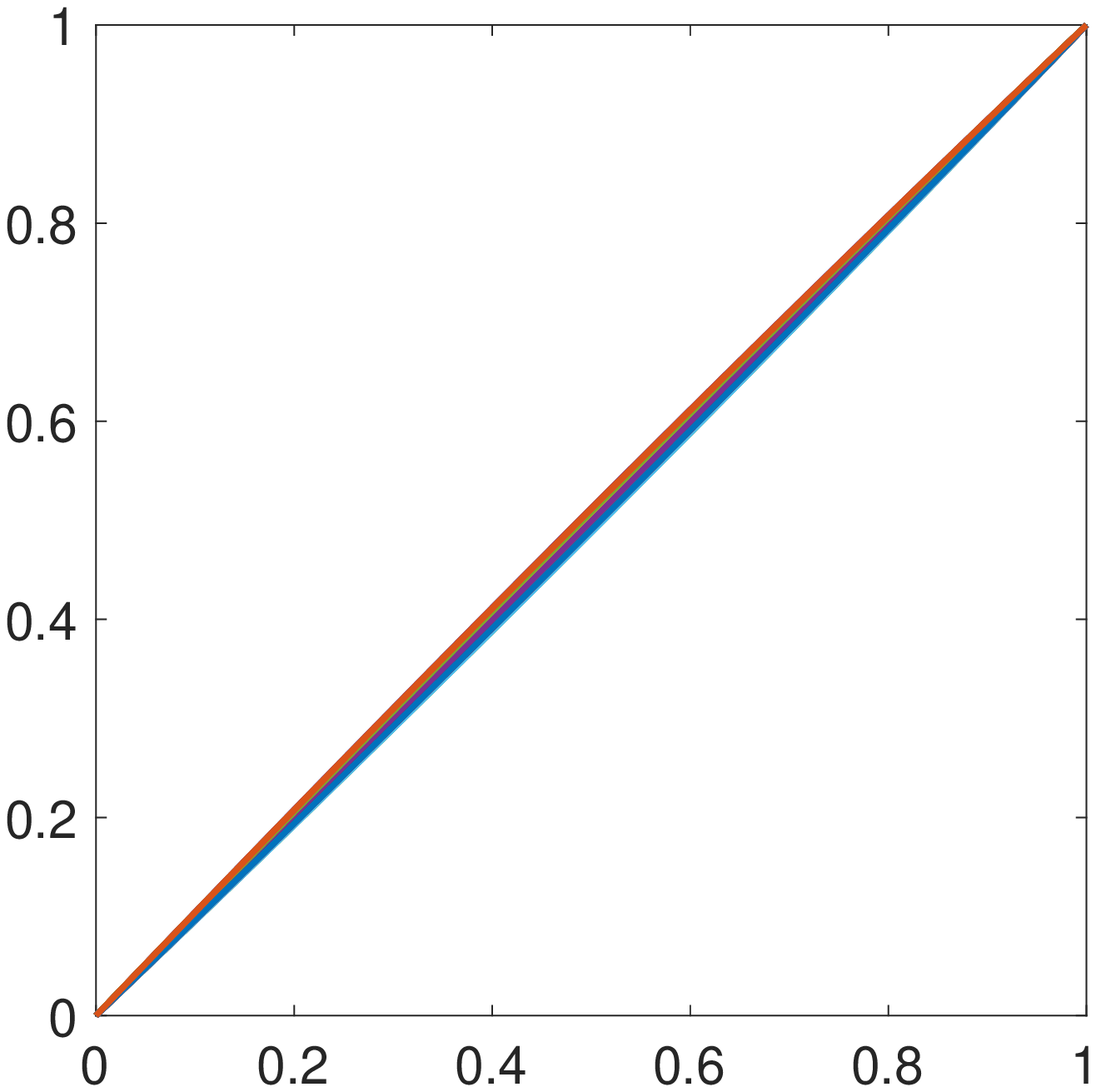} &
\includegraphics[height=1.1in]{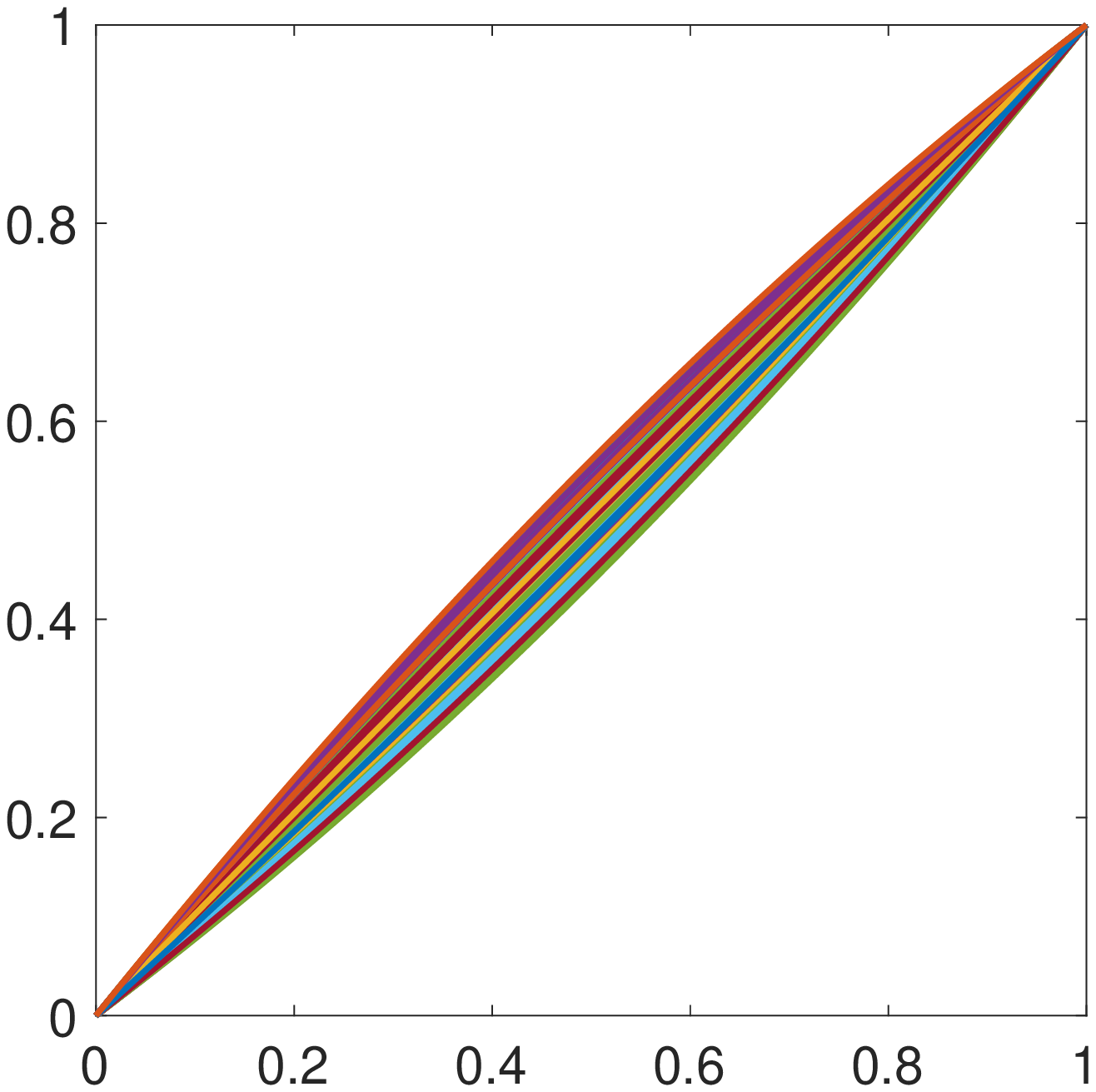} &
\includegraphics[height=1.1in]{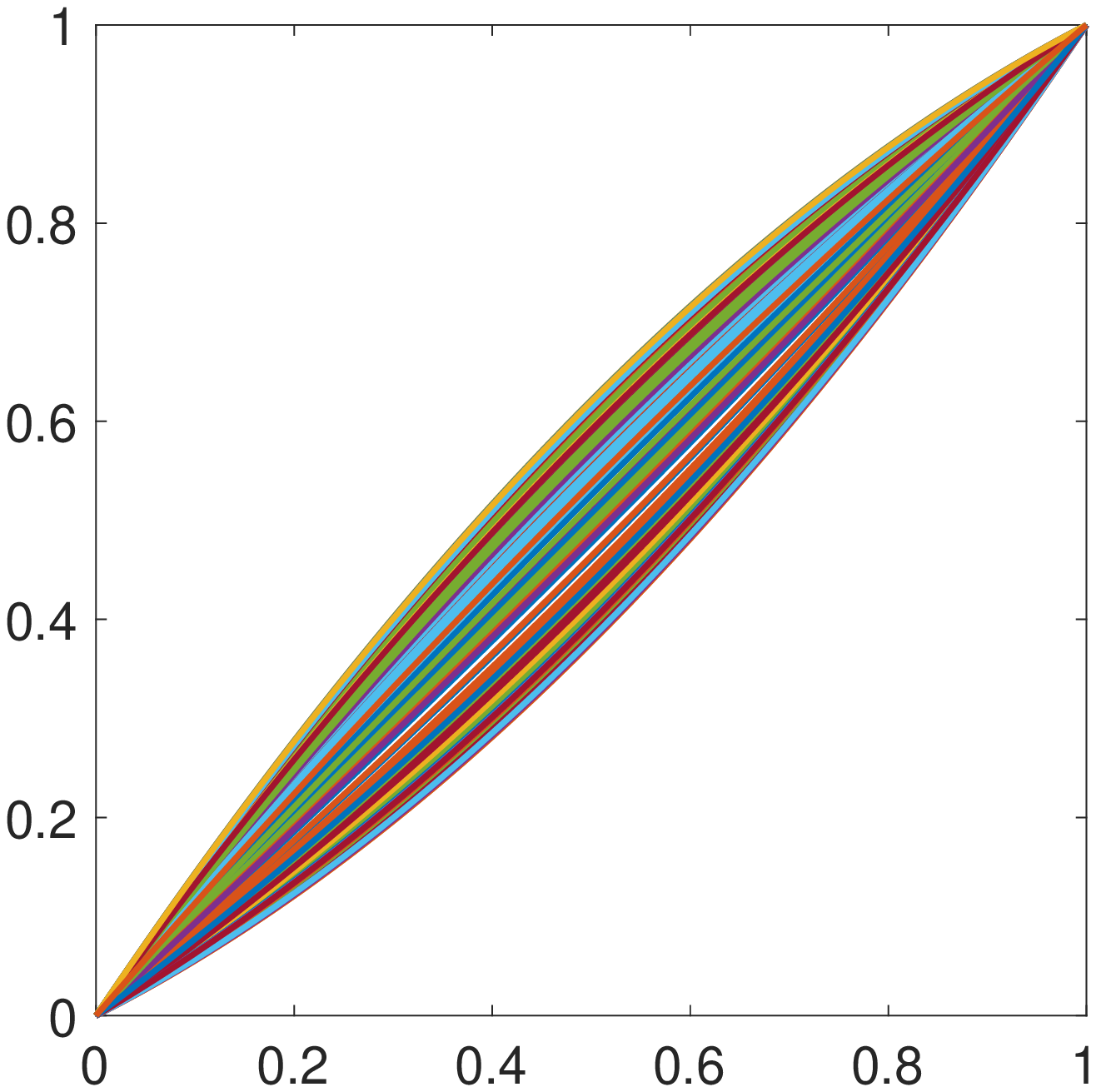} \\
Warping fns,  $a=0.05$ & Warping fns, $a = 0.25$ & Warping fns, $a = 0.5$   \\
\end{tabular}
\caption{Experiments show superlinear decrease in $R^2$ prediction measure
as the amount of phase variability is increased in predictor functions.} \label{fig:misregistration-demo}
\end{center}
\end{figure}

We further illustrate the issue of phase variability using a simulated example. 
Specifically, we quantify  deterioration in prediction performance as the 
amount of random warpings in the predictor functions is increased. The results are 
presented in Fig.~\ref{fig:misregistration-demo}.
The left panel shows a set of predictors $\{f_i\}$ used in these experiments. 
For a fixed $\beta$ and $\alpha = 0$, we
simulate responses $y_i$s using Eqn.~\ref{eq:regress_model}.
Then, we use this data $\{(f_i, y_i), i = 1,2,\dots,100\}$ to estimate the model parameters, including 
$\hat{\beta}$, using Eqn.~\ref{eq:ols}. Next, we use this estimated $\hat{\beta}$ to predict 
responses $y_i^{test}$ for new predictors $f_i^{test}$. However, let the test predictors be {\it contaminated}
in one of two ways: 
(i) value preserving $f_i^{test} \mapsto (f_i^{test} \circ \gamma_i)$, and (ii) 
area preserving $f_i^{test} \mapsto (f_i^{test} \circ \gamma_i) \sqrt{\dot{\gamma}_i}$.
Ignoring this contamination and using a standard predictor, we obtain predictions and quantify
prediction performance using the coefficient of determination $R^2$. 
Specifically, we study changes in $R^2$ 
as the amount of contamination (warping noise) increases. 
The warping functions used in this experiment are given by $\gamma_i(t) = t + \alpha_i t (1-t)$, 
where $\alpha_i \sim U(-a,a)$; the larger the value of $a$, the larger is the warping noise. 
The bottom row shows examples of warping functions for 
different values of $a$. 
The middle and the last panels in the top row show plots of $R^2$ versus $a$ (averaged over 200 runs) for the 
two types of contaminations. In both cases we observe a 
superlinear decay in the performance as $a$ increases. 
These experiments underline the fact that even a small amount of 
phase variability in predictors, either value-preserving or norm-preserving,  
can lead to a significant deterioration in the prediction performance. Thus, one needs to account for this 
variability inside the model itself in an intrinsic way. 

We reiterate that phase is nuisance in some but not all situations. 
One should not always expect the shapes of predictor functions to 
be predominant. Phase components may also carry important 
information about the responses and one should not 
always ignore them. However, in some cases, as illustrated through examples presented later in this paper, 
shapes are sometimes the primary predictors and one wants regression models that can exploit this knowledge.

\subsection{Potential Solutions}
This leads us to an important question: {\it What kind of regression models allow 
inclusion of only the shape or amplitude of the predictor 
functions and deemphasizes their phases}?  In general, there are some
parametric and nonparametric choices available.
\begin{enumerate}

\item {\bf Pre-Aligned Functional Linear Model (PAFLM)}: One 
obvious solution is to simply remove the phase variability in the given functions $\{f_i\}$ using one of 
several pre-existing functional alignment algorithms (see e.g. 
\cite{ramsay-li-RSSB:98,muller-JASA:2004,srivastava-etal-function:2011,tucker-etal:2013}). 
Then, one can 
use the aligned functions, or amplitudes,  for predicting the response variable using previously-mentioned FLM. 
The alignment algorithms are typically based on matching the 
given $\{f_i\}$ one-by-one to a template function which, in turn, 
is constructed iteratively using the means of the aligned functions. 
The limitation of this approach, in a regression setting, is that this alignment is performed
independent of the response variable. In other words, the values $\{ y_i\}$ do not play any role in 
the alignment.

\item {\bf Joint Modeling \& Alignment Under Value-Preserving Warping 
Using the $\ltwo$ Inner-Product}: Another possibility is to remove the 
phase within FLM by introducing an extra step.  
For instance, when using the contaminated predictors $\{ \tilde{f}_i = f_i \circ \gamma_i\}$, 
under the value-preserving
warping, one can try to
solve for the unknown warpings by adding optimization over $\gamma_i$s, as follows. 
We can modify the 
model in Eqn.~\ref{eq:regress_model} to become: 
\begin{equation}
	y_i = \alpha +  \sup_{\gamma_i \in \Gamma} \left( \int_0^T \tilde{f}_i(\gamma_i(t) )\beta(t)\,dt \right) +\epsilon_i,~~i=1,\dots,n\ .
		\label{eq:old_model}
\end{equation}
This additional optimization over $\Gamma$ is supposed to nullify the original contamination in $f_i$s. 
However, this approach as specified has a major shortcoming. As described in several places, 
see e.g. Marron et al. \cite{marron2014statistics} and Srivastava-Klassen \cite{srivastava2016functional}, 
the optimization over $\gamma_i$ 
under the $\ltwo$ inner product is actually degenerate, due to a phenomenon called the {\it pinching effect}. 
Some authors minimize pinching by restricting the set of warpings in Eqn.~\ref{eq:old_model} in a pre-determined manner. 
This restriction is unnatural as it is impossible to predict 
the optimal set of warpings needed to align future data.

\item {\bf Nonparametric Regression Model}:
Nonparametric models for functional regression 
are gaining popularity since they do not require any predetermined model and are purely data driven. 
Developed and studied by Ferraty and Vieu \cite{ferraty2006nonparametric}, 
a nonparametric model for functional regression is given by:
$y_i = G(f_i) + \epsilon_i$.
Here $G: {\cal F} \to \real$, an unknown smooth map,
and is estimated by the functional Nadaraya-Watson (NW) estimator \cite{nadaraya1964estimating}.
For the given data $\{(f_i, y_i),\ i = 1, 2, \dots, n\}$, the  estimator is given by:
\begin{equation}
\hat{G}(f) = {\sum_{i=1}^n y_i K(d(f_i, f)/b) \over \sum_{i=1}^n K(d(f_i, f)/b)},
\label{eq:kernel_cv}
\end{equation}
where
$K$ is the standard Gaussian kernel, $b$ is the bandwidth parameter, and 
$d$ is a distance on the predictor (function) space. 
Naturally, the choice of distance $d$ is critically important in such kernel estimators.  
If we use the standard $\ltwo$ norm in ${\cal F}$ for $d$, then the 
prediction will remain dependent on the phase of the predictors. 
Instead, if we choose a distance that compares {\it shapes of the predictors} and ignores their 
phases, {\it i.e.} $d$ is a proper
shape metric,  then the model 
becomes invariant to phase. 
\end{enumerate}

\subsection{Proposed Approach} 
There is possibility of a different parametric approach that stems from modifying
the main term in FLM (Eqn.~\ref{eq:regress_model})
directly, and making it invariant to the phase. 
This approach is motivated by the use of invariant metrics, such as
the Fisher-Rao metric and the elastic Riemannian metric in FDA \cite{srivastava-etal-function:2011,srivastava2016functional}. In fact, 
depending on the chosen warping, this elastic FDA 
framework gives several ideas although only a couple of them are discussed here. 
This framework is based on replacing the $\ltwo$ 
inner product and the $\ltwo$ distance in FDA 
by invariant Riemannian metrics and invariant distances
between functions. The invariant quantities provide better 
mathematical and numerical properties, and indeed lead to a superior 
performance in FDA. 
The challenge in using these
invariant metrics comes from their complicated expressions, but that is overcome using 
square root velocity function (SRVF) representation
(Srivastava et al. \cite{srivastava-etal-function:2011}). The SRVF of a function $f$ is
defined by: $q(t) = \mbox{sign}(\dot{f}(t)) \sqrt{| \dot{f}(t)|}$.  
One works with the SRVFs $q_i$s instead of the predictors $f_i$s 
and the Fisher-Rao metric simplifies to the standard $\ltwo$ metric under this change of variables. 
This framework motivates at least two ways of fixing the 
pinching problem in Eqn.~\ref{eq:old_model}: 

\begin{enumerate}
\item {\bf Use SRVF Representation and Value-Preserving Warping}: The first idea is to compute SRVFs of the 
given predictors, and then simply replace the term 
$\sup_{\gamma_i}  \inner{f_i \circ \gamma_i}{\beta}$ in Eqn.~\ref{eq:old_model} by the term: 
$\sup_{\gamma_i} \inner{(q_i \circ \gamma_i)\sqrt{\dot{\gamma}_i}}{\beta}$. 
This is motivated by the fact that under Fisher-Rao invariant metric, the inner product between functions is 
exactly equal to the $\ltwo$ inner product of their SRVFs.

The corresponding time warpings
of SRVFs, $q_i$s, are given by 
$(q_i \circ \gamma_i)\sqrt{\dot{\gamma}_i}$, and 
are norm preserving. 
That is, $\| q_i \| =  \|(q_i \circ \gamma_i)\sqrt{\dot{\gamma}_i}\|$ for 
all $q_i \in \ltwo$ and $\gamma_i \in \Gamma$, and thus pinching is no longer possible. 
More importantly, the model is now completely 
independent of the phase components of the predictors $f_i$s.

\item {\bf Use Original Functions and Norm-Preserving Warping}: The other option is to 
incorporate the norm-preserving transformations of the functions themselves 
($f_i \mapsto {(f_i \circ \gamma_i)\sqrt{\dot{\gamma}}_i}$) in the model, without 
resorting to SRVFs. 
As noted earlier, this warping changes both the locations and the heights of peaks 
and valleys in function, but preserves its
$\ltwo$ norm. 
In this case we replace the 
problematic $\ltwo$ inner-product term in Eqn.~\ref{eq:old_model} by the term 
$\sup_{\gamma_i}  \inner{(f_i \circ \gamma_i)\sqrt{\dot{\gamma}}_i}{\beta}$.
This option is especially
suitable when $f_i$s are noisy and an SRVF transformation may further enhance this noise. 
 By working with $f_i$s, one inherits all the nice properties of 
 the Fisher-Rao framework and avoids enhancing 
the noise. However, this warping is different from the value-preserving warping $f \circ \gamma_i$ used in 
traditional functional alignment.
Thus, $\gamma_i$s here can be called {\it phase} in a broader sense but not in a classical sense.
In the end, the regression model is invariant to the phase of the predictors, except the phase is now defined
using the mapping $f_i \mapsto (f_i \circ \gamma_i)\sqrt{\dot{\gamma}}_i$. 
\end{enumerate}
Each of these models help remove the phase variability, 
avoid the pinching effect, and improve prediction performance. Ultimately, the choice of a model
depends on the nature of the data and the goals of the application. 
The response variables in both these models are invariant
to the respective warpings of the predictor functions.  
In this paper, we will develop the second approach and will call this the 
{\it elastic functional regression} (EFRM) model.

The rest of this paper is as follows. 
In Section 2, we develop the proposed elastic functional regression model 
and discuss estimation of model parameters. We demonstrate this model using 
some simulated data and real data, and compare its performance against some current ideas
 in Section 3. Lastly, Section 4 ends the paper with some concluding remarks. 

\section{Elastic Functional Regression Model (EFRM)}
In this section, we layout a regression model for {\it scalar-on-function} problem
with the property that the response variable is invariant to the phase component of the predictor. 
This framework is based on ideas used previously for alignment of functional data, or phase-amplitude separation, 
using invariant metrics and the SRVF representation of functions. We start by briefly introducing those concepts 
and refer the reader to \cite{srivastava-etal-function:2011} for additional details. 

\subsection{Model Specification}

As mentioned earlier, the use of $\ltwo$ inner-product or $\ltwo$ norm for alignment of functions leads to a well-known problem 
called the {\it pinching effect}. While some papers avoid this problem using a combination of external penalties and search space
reductions, a more comprehensive solution comes from using an elastic Riemannian metric with appropriate invariance properties. This metric, 
called the {\it Fisher-Rao metric} for functions, avoids the pinching effect without any external constraints 
and results in superior alignment results. 
Let $f$ be a real-valued function on the interval $[0, 1]$ (with appropriate smoothness) and  let ${\cal F}$ denote the set of all such functions. Let $\Gamma$ be the set of all boundary preserving diffeomorphisms of the unit 
interval $[0,1]$, \ie  ~$\Gamma = \{ \gamma: [0,1] \to [0,1]\ |\ \gamma(0)= 0, \gamma(1) = 1, \gamma\ \mbox{is a 
diffeomoprhism}\}$. 
For the purpose of alignment, one represents a function $f$ using its square-root velocity function (SRVF):
$q(t) = \mbox{sign}(\dot{f}(t)) \sqrt{| \dot{f}(t) |}$. 
One of the advantages of using SRVF is that under the 
transformation $f \mapsto q$, the complicated Fisher-Rao Riemannian metric and the Fisher-Rao distance map
into much simpler expressions ($\ltwo$ inner product and $\ltwo$ norm, respectively).
If we warp a function $f$ by a time warping $\gamma$, \ie, map $f \mapsto (f \circ \gamma)$, then its SRVF changes
by $q \mapsto (q \circ \gamma) \sqrt{\dot{\gamma}}$. The latter is often denoted by $(q* \gamma)$. 
The invariance property of the Fisher-Rao metric implies that for any $q_1, q_2 \in \ltwo$ and $\gamma \in \Gamma$, 
we have:
$\| (q_1* \gamma) - (q_2* \gamma) \| = \| q_1 - q_2 \|$. 
In other words, the action of $\Gamma$ on $\ltwo$ is by isometries. A special case of this equation is that 
$\|(q* \gamma)\| = \| q\|$ for all $q$ and $\gamma$. Thus, this action preserves the $\ltwo$ norm of 
the SRVF and, therefore, avoids any pinching effect.

This framework motivates several solutions for avoiding the pinching problem associated with the 
inner-product term in Eqn.~\ref{eq:old_model}. While one can work with the SRVFs of the given predictor functions, 
they are prone to noise in the original data due to the involvement of a time derivative
in the definition of SRVF. In case the original data is noisy, this noise
gets enhanced by taking a derivative. As a workaround to this problem, we treat the given predictor functions to 
be in the SRVF space already. That is, we assume the action of warping $\gamma_i$ on an $f_i$s is given by 
$ (f_i \circ \gamma_i) \sqrt{\dot{\gamma}_i}$ and not $f_i \circ \gamma_i$. With this action, we have that 
$\| (f_i* \gamma_i)\| = \| (f_i \circ \gamma_i) \sqrt{\dot{\gamma}_i}\| = \| f_i\|$. 

Based on this argument, the inner-product term in Eqn.~\ref{eq:old_model} can be replaced by the term: 
$\sup_{\gamma_i \in \Gamma} \inner{\beta}{(f_i* \gamma_i)}$. 
This is a scalar quantity and represents a modified linear relationship between the predictor and the response. 
One can impose a single-index model on top of this construction to generalize this model. 
Such single-index models have been used commonly in conjunction with 
FLMs, see e.g. \cite{stoker1986, ait2008, reiss2017, eilers, jiang2011functional}.
For any $h: \real \to \real$, a smooth function, define EFRM the model: 
\begin{equation}
y_i = h \Big( \sup_{\gamma_i \in \Gamma} \inner{\beta}{(f_i* \gamma_i)}  \Big) + \epsilon_i, i = 1, \dots , n
\label{eqn:full-model}
\end{equation}
To complete model specification, we assume $\epsilon_i$s to be ${\it i.i.d}$ zero-mean,  Gaussian random 
variables. 
This model has the following properties. 
\begin{enumerate}
\item {\bf Nonlinear Relationships}: There are two sources of nonlinearity in the 
relationship between $f_i$ and $y_i$. Although the inner product $\inner{\beta}{f_i}$ is linear in $f_i$, 
the supremum over $\Gamma$ makes the term $\sup_{\gamma_i \in \Gamma} \inner{\beta}{(f_i* \gamma_i)}$
nonlinear. Furthermore, the inclusion of $h$ allows EFRM to makes relationship firmly nonlinear.
 
\item {\bf Invariance to Phase}: For a fixed model description $(\beta, h)$, the mean of
response $y_i$  is invariant to 
the phase of $f_i$ due to the fact that 
$\sup_{\gamma_i} \inner{\beta}{(f_i* \gamma_i)} = \sup_{\gamma_i} \inner{\beta}{((f_i* \gamma_0)* \gamma_i)}$, 
for all $\gamma_0 \in \Gamma$. Even though the mean of $y_i$ is invariant to the phase, we note that 
the estimated values of $\beta$ and $h$ (covered in the next section) can depend on the 
phase of $f_i$. 

\item {\bf Identifiability of $\beta$}: In view of the equality mentioned in the previous item, 
the regression coefficient $\beta$
is not fully specified. This is because if $\hat{\beta}$ is an estimator of $\beta$, then so is 
$\hat{\beta} \circ \gamma_0$ for 
any $\gamma_0 \in \Gamma$. To avoid this ambiguity, we impose an additional constraint on the model that all 
the maximizers $\{\hat{\gamma}_i 
 = \arg\sup_{\gamma_i} \inner{\beta}{(f_i* \gamma_i)}\}$ together 
 satisfy the condition that ${1 \over n}\sum_{i=1}^n \hat{\gamma}_i = \gamma_{id}$. 

\item {\bf Difference from GFLM}: The single-index model used here is 
quite similar to a generalized FLM (GFLM), but with an important difference. 
In a single-index model, the index function $h$ is unknown and needs to be estimated from the data itself, while in 
generalized model $h$ is assumed known.  
One can easily switch from EFRM to GFLM, if needed, by using a known $h$. 

\end{enumerate}

\subsection{Parameter Estimation}

Next we consider the problem of estimating EFRM parameters using MLE. The unknown parameters are: 
the 
index function $h$ and
the coefficient of regression $\beta$. We take an iterative approach, 
laid out in  \cite{eilers2009multivariate}, where one updates
estimates of $h$ or $\beta$ while keeping the other fixed. 
Thus, we first focus on techniques for estimating 
these quantities separately.
 
 \paragraph{\bf Estimation of $\beta$ Keeping $h$ Fixed}:
Given a set of observations $\{ (f_i, y_i )\}$, the
goal here is to solve for MLE of $\beta$, 
while keeping $h$ fixed. In order to reduce the search space to a finite-dimensional set, 
we will assume that $\beta \in \{  \sum_{j=1}^J c_j b_j | c_j \in \real\}$ for a fixed 
orthonormal basis ${\cal B} = \{ b_j, j=1,2,\dots\}$ of 
$\ltwo([0,1],\real)$. 
The estimation problem is now given by: 
\begin{eqnarray*}
\hat{c} = \argmin_{c \in \real^J} \ \ H(c),\ \ \mbox{where}\ \ H:\real^J \to \real, \ \mbox{given by}\\ 
H(c) = \left( \sum_{i=1}^n (y_i - h( \sup_{\gamma_i \in \Gamma} \inner{\sum_{j=1}^J c_j b_j}{(f_i* \gamma_i)})^2 \right)\ .
\end{eqnarray*}
We use a MATLAB function {\tt fminunc}, which use the 
quasi-Newton method, to solve the minimization problem.
Contained within this problem are a set of optimizations over $\gamma_i$s. 
For a fixed $c$, this optimization is performed using the dynamic programming algorithm (DPA) for 
each $i=1,2,\dots, n$. This set of calls to DPA are inside the definition of $H$ and are performed for 
each candidate value of $c$. 
Thus, any update of $c$ requires recomputing the optimal warping functions, 
resulting in an iterative process. 
Finally, once $c$ (or $\beta$) is estimated, we can impose the condition for specification of 
$\beta$, {\it i.e.}
${1 \over n}\sum_{i=1}^n \hat{\gamma}_i = \gamma_{id}$ as follows. 
For this, we use the current $\hat{\gamma}_i$s to compute their 
average $\bar{\gamma} = {1 \over n}\sum_{i=1}^n \hat{\gamma}_i$ and replace $\beta$ by $\beta \circ \bar{\gamma}^{-1}$. 
The full process for estimating $\beta$ is summarized in Algorithm 1.

\begin{algorithm}[htb]
    \caption{Estimation of $\beta$ keeping $h$ fixed}
    \label{algo:beta}
    \begin{algorithmic}[1]
        \State  Initialization Step. Choose an initial $c \in \real^J$ and compute $\hat{\beta}(t) = \sum_{j=1}^{J}c_j b_j(t)$.
        \State Use an optimization method (such as {\tt fminunc} in MATLAB) to find $\hat{c}$
       that minimizes the cost function $H$.

\begin{itemize}
        \State To define $H$, use the current $\hat{c}$ (and $\hat{\beta}$) to perform the following  for each $i=1,2,\dots, n$,  
        \begin{itemize}
        \item Solve for $\hat{\gamma}_i = \argmin_{\gamma \in \Gamma} \| \hat{\beta} - (f_i* \gamma_i)\|^2$, 
        using the Dynamic Programming algorithm 
        (DPA).
\item  Compute the aligned functions $\tilde{f}_i \leftarrow (f_i * \gamma_i ) \equiv (f_i \circ \hat{\gamma}_i) \sqrt{\dot{\hat{\gamma}}_i}$.
\end{itemize}
\end{itemize}
                \State Update $\hat{\beta}(t) = \sum_{j=1}^J \hat{c}_j b_j(t)$. If the $|H(\hat{c})|$ is large, then return to step 2.
        \State Compute $\bar{\gamma} = {1 \over n}\sum_{i=1}^n \hat{\gamma}_i$ and replace $\beta$ by $\beta \circ \bar{\gamma}^{-1}$.
    \end{algorithmic}
\end{algorithm}

To analyze this estimator, one has to study the choice of $J$ relative to the sample size $n$, and develop an 
asymptotic theory. Since this analysis is very similar to existing papers involving 
functional predictors \cite{li2010generalized, morris2015functional}, we simply refer to that literature for asymptotic analysis. 

\paragraph{\bf Estimation of $h$ Keeping $\beta$ Fixed} 

Next we consider the problem of estimating the index function $h$ given the data and the current
estimate of $\beta$.  The reason for introducing this single-index model is to capture nonlinear relationship between 
the predicted responses and observed responses. While there are many potential nonparametric estimators for $h$, 
we keep the model simple by restricting to lower-order polynomials. 
We allow $h$ to be only linear, quadratic, and cubic:  
$h(x) = ax +b$, $h(x) = ax^2 + bx + c$, and $h(x) = ax^3 + bx^2 + cx + d$, etc. 

For estimating $h$, we first predict responses according to: 
$ \hat{y}_i  = \sup_{\gamma_i \in \Gamma} \inner{\hat{\beta}}{(f_i* \gamma_i)}$,
and then we fit a polynomial function $h$ between the predicted responses 
 $\hat{y}_i$ and the observed responses 
$y_{i}$ using the least squares error criterion.
The full parameter estimation procedure is presented in 
Algorithm \ref{algo:final}. 
\begin{algorithm}[htb]
    \caption{Elastic Functional Regression Model}
    \label{algo:final}
    \begin{algorithmic}[1]
        \State  Initialize $h$ as the identity function ($h(x) = x$). 
	\State Given $h$, use Algorithm \ref{algo:beta} to estimate $\hat{\beta}$.
        \State For a given $\hat{\beta}$, update $h$ using the least squares criterion.
\State If $|H(\hat{c})|$ is small, then stop. Else, return to step 2.
    \end{algorithmic}
\end{algorithm}

\subsection{Prediction of Response Under the Elastic Regression Model}
One of the goals of EFRM is to predict values of the response variable for
the future predictor observations. Here we describe the prediction process under EFRM. 
As the model suggests, this prediction is based on alignment of predictors to the coefficient  
$\hat{\beta} = \sum_{j=1}^J \hat{c}_j b_j$
using DPA. For a given predictor $f^{(test)}$, the predicted 
value of $y$ is:
\begin{equation}
\hat{y}^{(test)}= \hat{h} \Big( \sup_{\gamma_i \in \Gamma} \inner{\sum_{j=1}^J \hat{c}_j b_j}{(f^{(test)} * \gamma_i)} \Big).
\label{eqn:prediction}
\end{equation}
We will use this predictor to evaluate prediction performance of EFRM, relative to
current models, using both  simulated data and real data.

\section{Experimental Illustration}

We will compare EFRM with four natural alternatives.
Either these models are commonly used in the literature or they are simple modifications
of the current models for handling the phase variability in the predictors. 
These models are:  
Functional Linear Model (FLM); Pre-Aligned Functional Linear Model (PAFLM); Nonparametric regression model (NP)  using a Gaussian kernel function and 
two different choices of $d$. 
We briefly summarize and introduce these models.

\paragraph{\bf Functional Linear Model (FLM)}
FLM has already been introduced 
in Eqn.~\ref{eq:regress_model}. As stated earlier, it does not specifically account for the presence of 
phase variability in the predictor data and is vulnerable to that nuisance variability.

\paragraph{\bf Pre-Aligned Functional Linear Model (PAFLM)}
PAFLM is the model where one pre-aligns the predictor functions (using a phase-amplitude separation algorithm)
and then performs standard FLM. To clarify further, one performs phase-amplitude separation and then 
discards the phase component. 
In the results presented here, we use the {\it ``Complete Alignment Algorithm''}
presented in \cite{srivastava-etal-function:2011}. 
This alignment is suboptimal from the perspective of regression, since the response variable is 
not used in the alignment. 

\paragraph{\bf Nonparametric Kernel Approach}
As mentioned earlier, one can use the Nadaraya-Watson estimator (of the kind given in Eqn.~\ref{eq:kernel_cv}) for 
predicting $y$ for a new predictor function $f$. The only quantity left unspecified in that equation is the metric structure on ${\cal F}$. 
In the following we choose the distance to be either the $\ltwo$ norm or a weight shape distance. 
The weighted shape distance uses a pre-alignment of predictor functions and is defined as follows. 
Let the predictors $\{f_i\}$ be pre-aligned (as discussed above) resulting the phases $\{\hat{\gamma}_i\}$
and amplitude $\{ f_i * \hat{\gamma}_i\}$. Then, define the distance
$d(f, f_i) = \lambda d_a(f,f_i) + (1 - \lambda) d_p(f,f_i)$, where $\lambda \in [0, 1]$ is a proportion parameter. 
Here $d_a$ denotes the amplitude distance: 
$d_a(f,f_i) = \|f - (f_i * \hat{\gamma}_i)\|$
and $d_p$ denotes the phase distance: $d_p(f,f_i) = \|\sqrt{\dot{\hat{\gamma}}_i} - \sqrt{\dot{\gamma}_{id}}\|$.
The optimal value of the bandwidth
$\hat{b}$ can be obtained via cross-validation:
$$\hat{b} = \argmin_{b \in \real_+} \sum_{i=1}^n (y_i - G_{(-i)} (f_i))^2,\ \ \ 
\mbox{with}\ \ \ 
G_{(-i)} (f) = \frac{\sum_{j=1, j \neq i}^{n} y_j K(d(f_j, f))/b)}{\sum_{j=1, j \neq i}^{n} K(d(f_j, f))/b)}
$$
For the joint estimation of $\lambda$ and $b$, we first compute the optimal bandwidth $\hat{b}$ for each $\lambda \in [0, 1]$. 
Then, we choose the optimal $\hat{\lambda}$ which gives the lowest cross-validation error.

Next, we present experimental results from these and EFRM on a number of data sets.

\subsection{Simulation Study}

In the studies presented in this section, we perform a five-fold cross-validation and compute the mean and standard deviation of root mean square error (RMSE) for predicting the response variable. We use this RMSE for comparing performances of different regression models.\\

\subsubsection{Simulated Data 1}

\begin{figure}[!htb]
    \centering
    \begin{subfigure}[!htb]{.45\textwidth}
        \includegraphics[width=\textwidth]{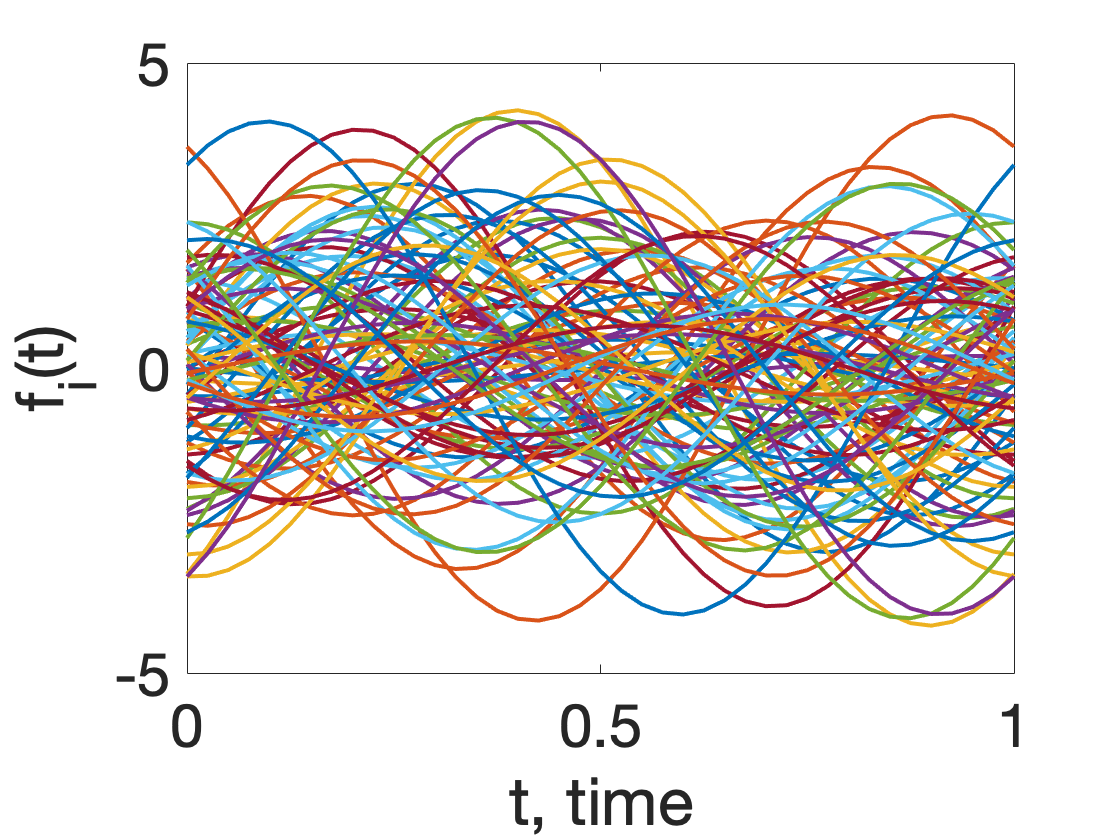}
        \caption{$\{ f^0_i \}$}
        \label{fig:simulated_f}
    \end{subfigure}
    ~ %add desired spacing between images, e. g. ~, \quad, \qquad, \hfill etc. 
    \begin{subfigure}[!htb]{.45\textwidth}
        \includegraphics[width=\textwidth]{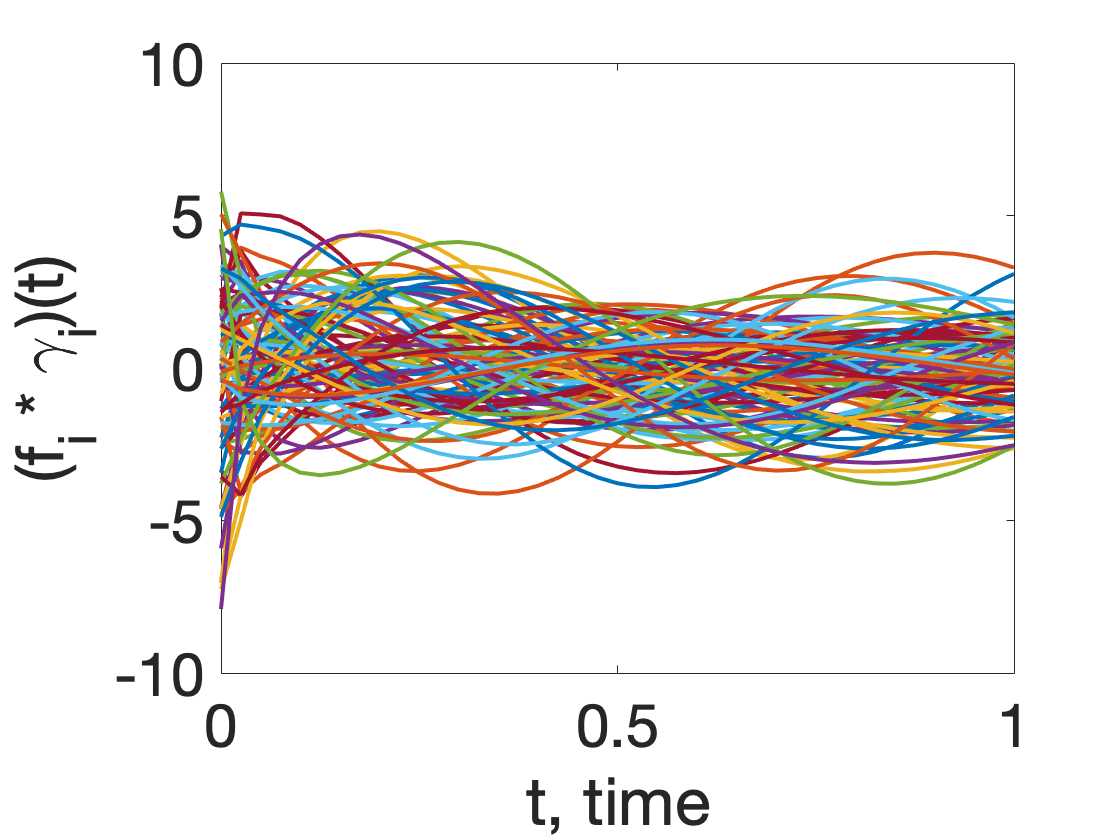}
        \caption{$\{( f_i = f^0_i *\gamma_i )\}$}
        \label{fig:simulated_warpedf}
    \end{subfigure}
        \begin{subfigure}[!htb]{.45\textwidth}
        \includegraphics[width=\textwidth]{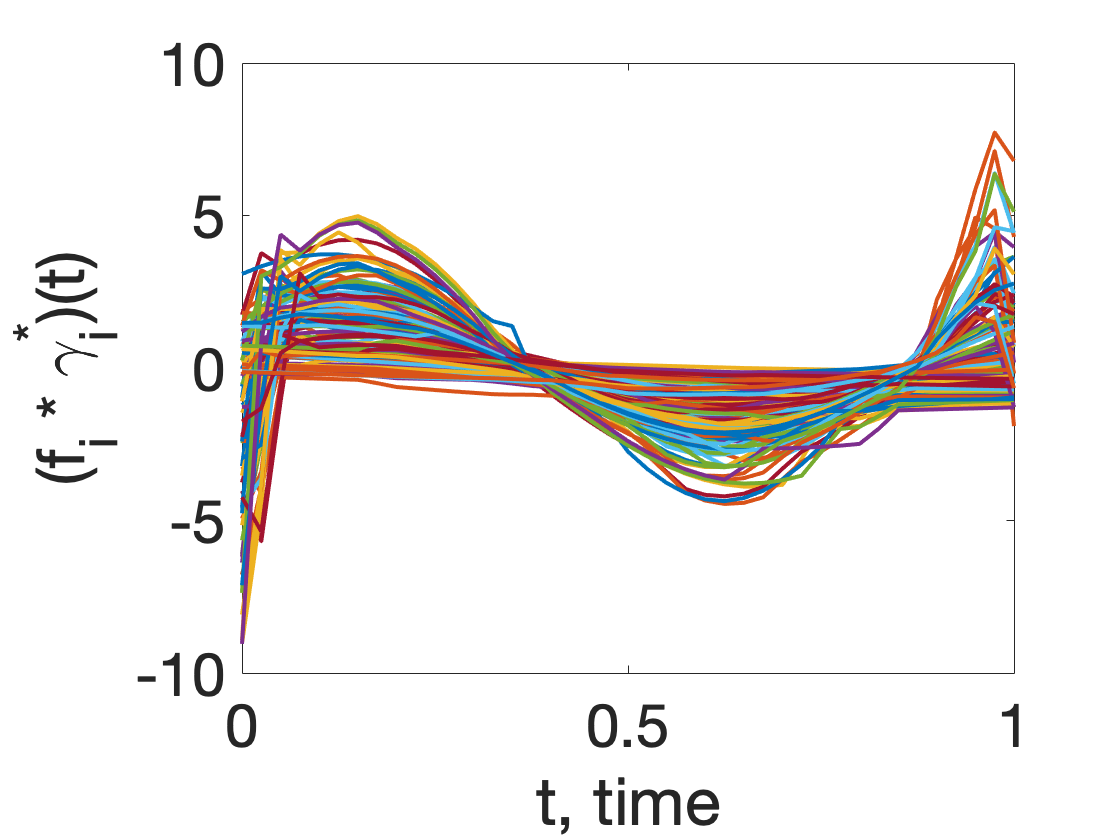}
        \caption{$\{( f_i * \gamma^*_i )\}$}
        \label{fig:simulated_warpedf}
    \end{subfigure}
        \begin{subfigure}[!htb]{.45\textwidth}
        \includegraphics[width=\textwidth]{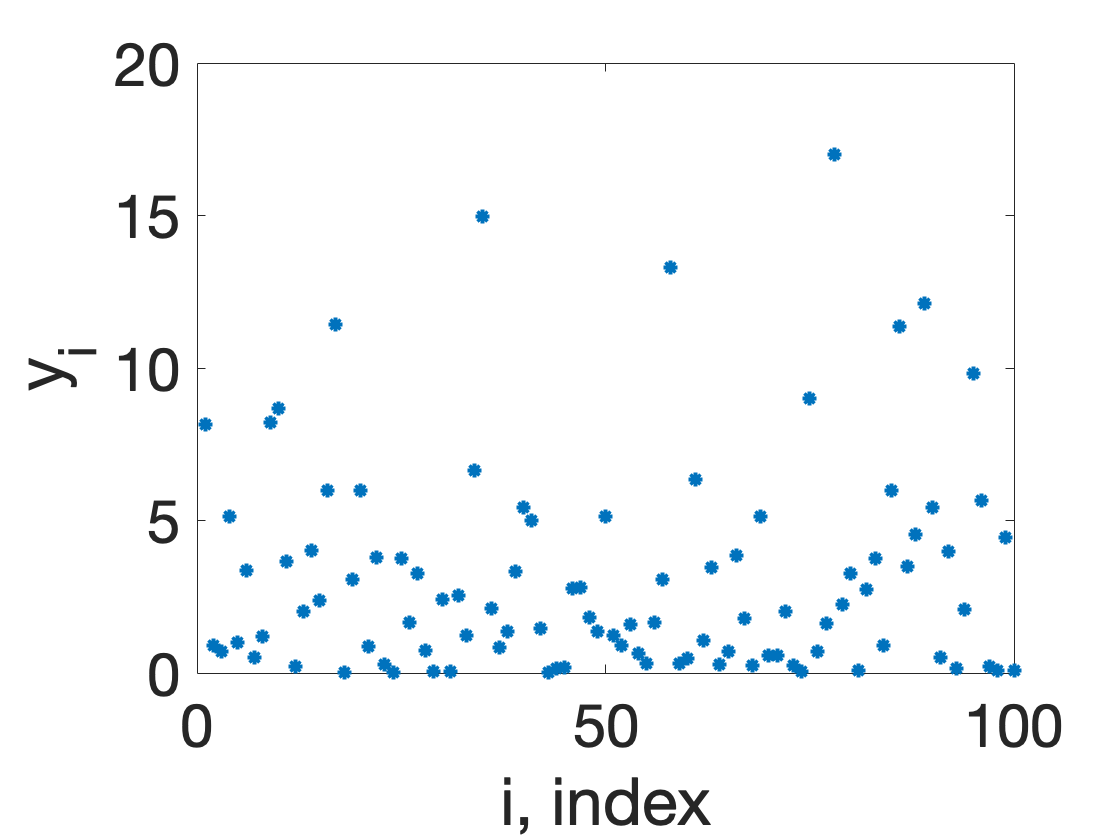}
        \caption{$\{y_i\}$}
        \label{fig:simulated_y}
    \end{subfigure}
    \caption{Simulated data 1. (a) shows the original functions, $\{f^0_i\}$, (b) shows them after
    random warpings, $\{ f_i\}$, (c) shows predictors after optimizations over 
    $\gamma_i$ in the generative model in Eqn.~\ref{eqn:full-model}, $\{f_i* \gamma^*_i\}$, 
    and (d) displays ordered response variables, $\{y_i\}$, from that model.}
    \label{fig:simulated_estimation}
\end{figure}

In the first experiment, we simulate $n = 100$ observations using 
the model stated in Eqn.~\ref{eqn:full-model}. For the
predictors, we use a Fourier basis and 
random coefficients to form the functions, $f^0_i(t) = c_{i,1} \sqrt{2} sin(2 \pi t) + c_{i,2} \sqrt{2} cos (2 \pi t)$ 
with $c_{i, 1}, c_{i, 2} \sim N(0, 1^2)$. 
Given these functions, we 
perturb them using random $\{\gamma_i\}$ to obtain the predictors $\{ f_i = (f^0_i*\gamma_i)\}$. 
We also simulate the coefficient function $\beta$ using the same Fourier 
basis but with a fixed coefficient vector  $c_0 = [1, 1]$. We plug these quantities in the 
model, use a quadratic polynomial for $h$, and 
add independent observation noise, $\epsilon_i \sim N(0, 0.01^2)$, to obtain responses $\{y_i\}$. 
This process is illustrated in Fig.~\ref{fig:simulated_estimation}.
We use a random 80-20 split for training and testing, respectively.

\paragraph{\bf Model Estimation}
Using the training data, we estimate the model parameters $h$ and $\beta$, 
as described in Algorithm \ref{algo:final}.  
In order to evaluate this algorithm, we 
 use three different bases for estimating $\beta$ during training: 1) Fourier basis with 
 only two elements, 2) Fourier basis with four elements, and 3) B-spline basis with four elements. 
The reason for using different bases for estimation is to 
study the effects of basis on the model performance. 
We also try three different polynomials: linear, quadratic, and cubic, as $h$ during estimation.

\begin{figure}[!htb]
    \centering
    \begin{subfigure}[b]{0.45\textwidth}
        \includegraphics[width=\textwidth]{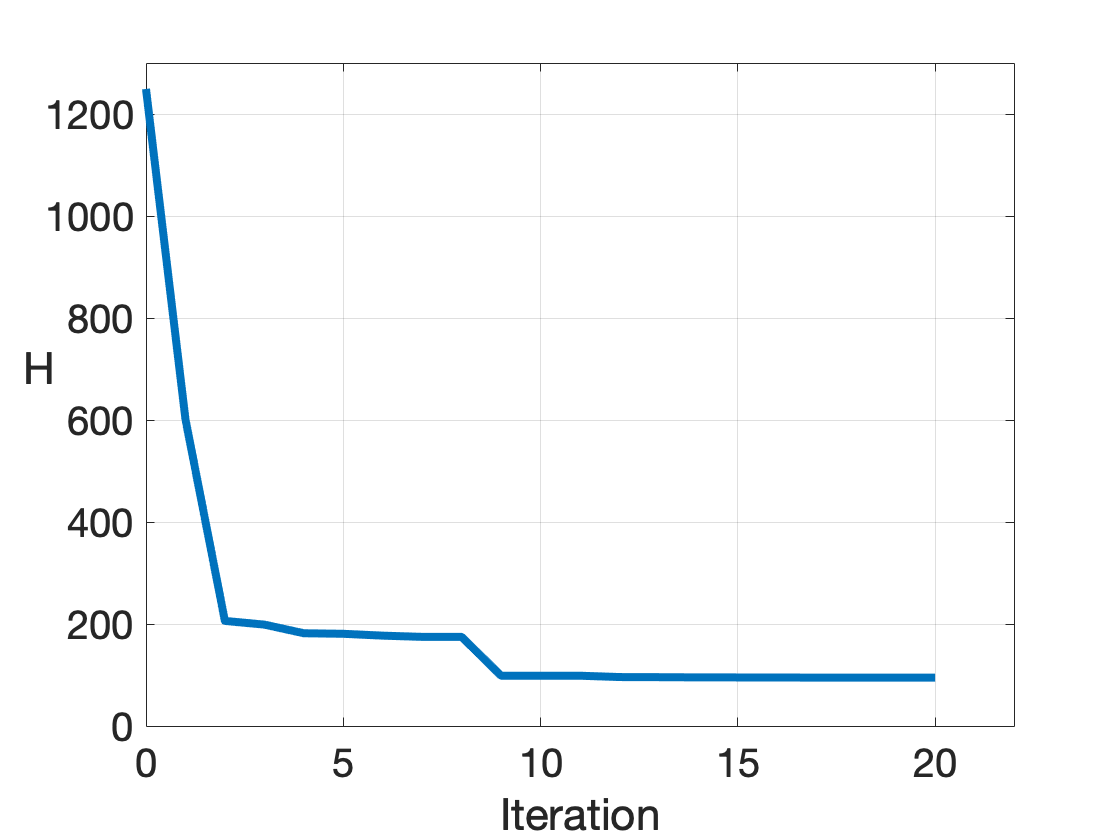} 
       \caption{$h$: linear}
        \label{fig:6}
    \end{subfigure}
    ~ %add desired spacing between images, e. g. ~, \quad, \qquad, \hfill etc. 
    \begin{subfigure}[b]{0.45\textwidth}
        \includegraphics[width=\textwidth]{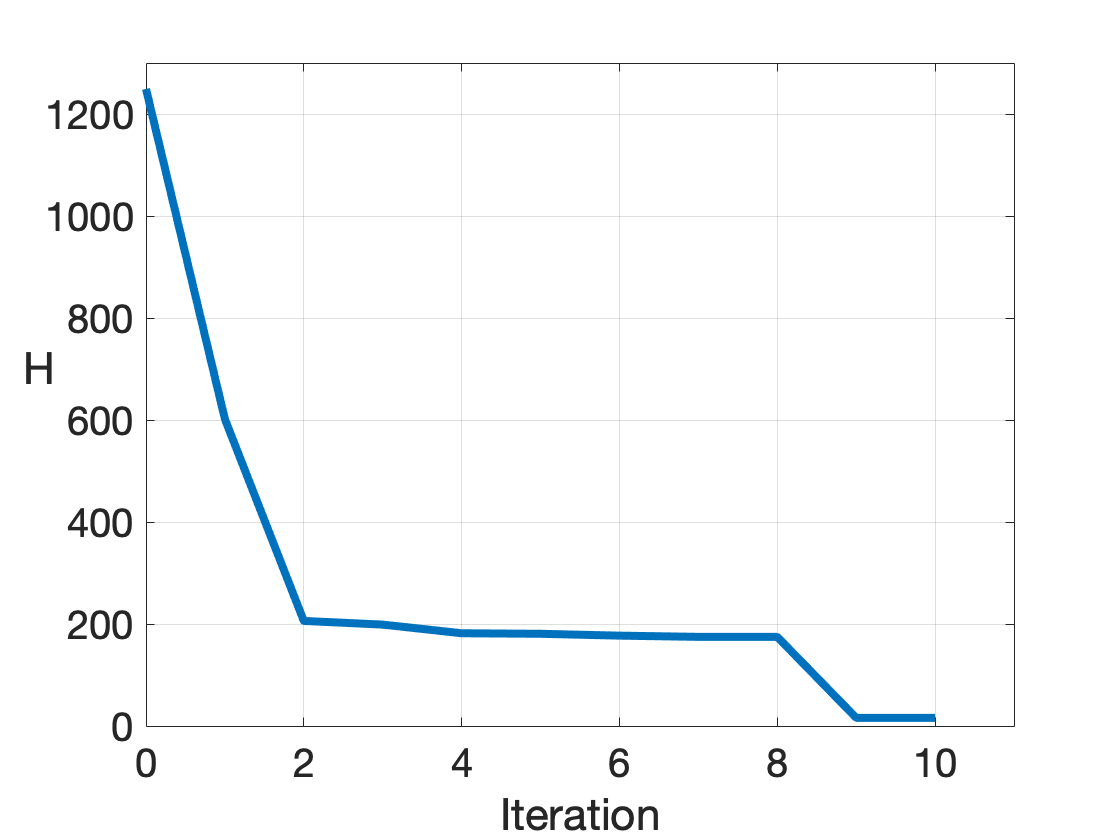} 
        \caption{$h$: quadratic}
        \label{fig:7}
    \end{subfigure}
        \begin{subfigure}[b]{0.45\textwidth}
        \includegraphics[width=\textwidth]{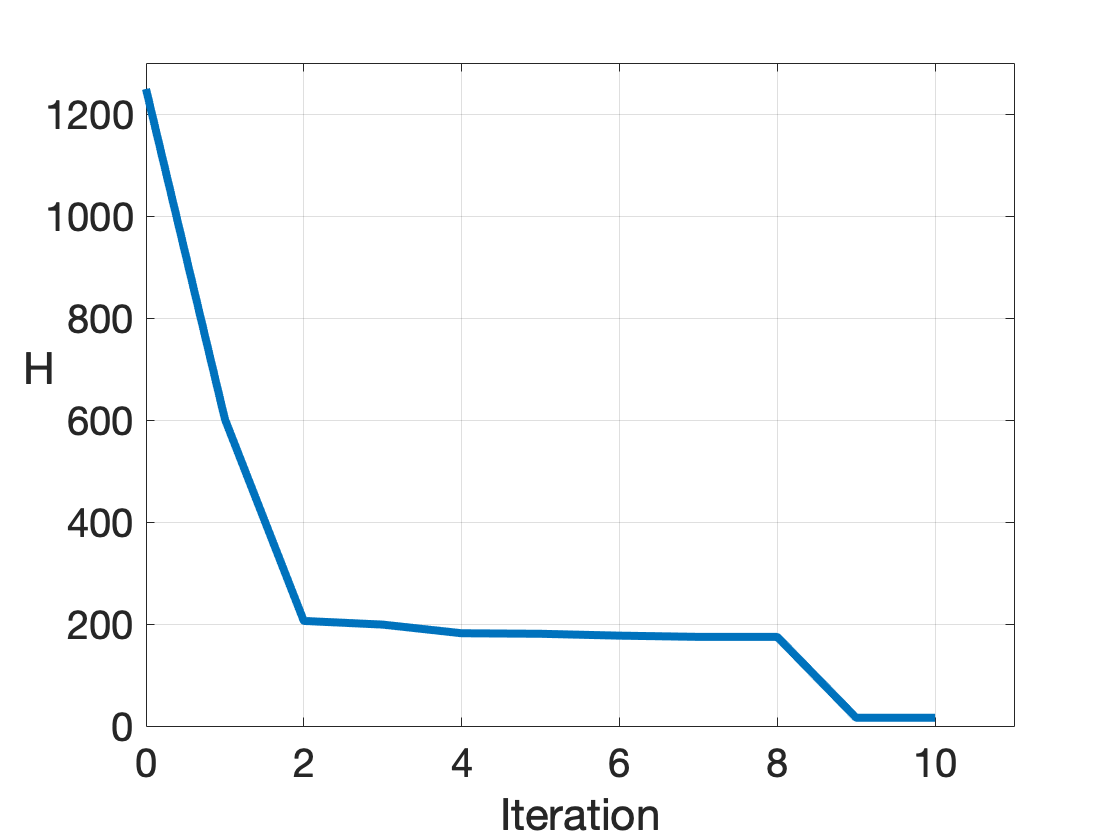} 
       \caption{$h$: cubic}
        \label{fig:8}
    \end{subfigure}
    \caption{The evolution of cost $H$ for each choice of the index function, $h$, and using Fourier basis with two elements
    for $\beta$.}
    \label{fig:simulated_estimation2}
\end{figure}

Fig. \ref{fig:simulated_estimation2} shows the evolution of cost function $H$ during optimization  in
Algorithm \ref{algo:final} for each of index functions: linear, quadratic, and cubic, 
in Fig. \ref{fig:6}, \ref{fig:7}, and \ref{fig:8}, respectively. These experiments use a Fourier basis with two elements to estimate $\beta$. 
These plots show that the cost $H$ goes down in all cases and the optimization algorithm provides 
at least local solutions reliably. The optimized values are found to be the best for the quadratic and 
cubic $\hat{h}$, which makes sense since 
a quadratic $h$ was used to simulate the data.

It is also important to quantify estimation performance for model parameters 
$\beta$ and $h$. In order to quantify 
these errors, we calculate the Root Squared Error 
RSE$_{\ltwo} = \sqrt{\int [a(t) - \hat{a}(t) ]^2 ~dt}$, where $a(t)$ is a functional parameter and $\hat{a}(t)$ is 
its estimate (for $a = \beta,\ h$). 
We then compute the averages of RSE$_{\ltwo}$ over a five-fold cross-validation.
The estimation errors for $\beta$ and $h$ for this simulation experiment
are presented in Table \ref{table:simulated_estimation}. Overall, the choice of a cubic 
$h$ does well in the estimation. 
If we compare these RSEs with prediction performances in Table \ref{table:simulated}, 
we see that a better estimation of $\beta$ and $h$ provides a better predictor 
of the response variable, which is natural.

\begin{table}[thb!]
\centering
\begin{tabular}{||c||cc|cc|cc||}
\hline \hline
 Basis              & \multicolumn{2}{c|}{Fourier2}   & \multicolumn{2}{c|}{Fourier4}   & \multicolumn{2}{c||}{Bspline4}   \\ \hline
Parameter& $\beta$        & $h$            & $\beta$        & $h$            & $\beta$        & $h$            \\ \hline
\hline
$h$: Linear    & 2.326          & 1.372         & \textbf{2.726} & 1.077          & 9.607         & 1.075          \\
$h$: Quadratic & {\bf 2.268} & 0.284 & 2.862          & 0.247          & 9.914          & 0.278          \\
$h$: Cubic     & 2.288          & \textbf{0.283} & 2.777          & \textbf{0.231} & \textbf{8.803} & \textbf{0.276} \\ \hline \hline
\end{tabular}
\caption{The average of RSE$_{\ltwo}$ (Root Squared Error) of $\hat{\beta}$ and $\hat{h}$ for
different choices of parameter sets on simulated data 1.}
\label{table:simulated_estimation}
\end{table}

%\begin{table}[thb!]
%\centering
%\begin{tabular}{||c|c|c|c||}
%\hline \hline
%\textbf{Basis}             & \textbf{Model} & \textbf{RSE$_{\ltwo}$ for $\beta$} & \textbf{RMSE$_{\ltwo}$ for $\gamma$} \\ \hline \hline
%\multirow{2}{*}{Fourier2}  & EFRM            & {\bf 68.636 }    & {\bf 0.017  }     \\ \cline{2-4} 
%                           & PAFLM          & 93.728                          & 0.025                                 \\ \hline
%\multirow{2}{*}{Fourier4}  & EFRM            &  {\bf 71.567}                           & {\bf  0.018 }   \\ \cline{2-4} 
%                           & PAFLM          & 209.831                          & 0.025                               \\ \hline
%\multirow{2}{*}{Bspline4} & EFRM            & {\bf 91.810}                           & {\bf 0.012}                               \\ \cline{2-4} 
%                           & PAFLM          & 119.430                           & 0.015                              \\ \hline \hline
%\end{tabular}
%\caption{The average of RSE$_{\ltwo}$ (Root Squared Error) of $\hat{\beta}$ and the RMSE$_{\ltwo}$ 
%$\{ \hat{\gamma}_i\}$ for different basis functions.}
%\label{table:simulated_estimation}
%\end{table}

\paragraph{\bf Prediction Performance}
To evaluate prediction performance, we use the model parameters estimated using the training step for 
predicting the response variable for the test data. This prediction follows the procedure 
laid out in Eqn.~\ref{eqn:prediction}. The predicted responses are then compared with the 
true responses to quantity the prediction error. 
We perform five-fold cross-validation to evaluate this error more precisely. 
Then we compute the average and the standard deviation of RMSE ($\sqrt{{1 \over n} \sum_{i=1}^n (y_i - \hat{y}_i)^2}$) from 
five different folds and use these quantities to compare different models.

The  results for average five-fold RMSEs and corresponding standard deviations 
are shown in Table \ref{table:simulated}.
As these results show,  EFRM is able to provide a 
better prediction performance than the competing models despite using very simple
tools. The predictions from PAFLM are less accurate since this method pre-aligns 
functional predictors without considering response variables $\{y_i\}$. The nonparametric regression model using 
the $\ltwo$ norm shows some improvement in prediction, when compared to FLM and PAFLM, 
since it is not restricted to linear relationships between the response and functional predictors. However, this model also fail to 
account for the phase variations and the predictions are found to be less accurate than EFRM.

\begin{table}[ht!b]
\centering
\begin{tabular}{||c||c|c|c||}
\hline \hline
\multicolumn{4}{||c||}{\bf Parametric} \\
\hline
~~~~~~Basis~~~~~~ & ~~~~ Fourier2 ~~~~& ~~~~ Fourier4 ~~~~& ~~~~ Bspline4 ~~~~\\ \hline \hline
$h$: Linear      & 1.140 (0.130)& 1.109 (0.257) &  1.604 (0.270)\\ 
$h$: Quadratic   &  0.527 (0.308) & 0.599 (0.213) & 1.509 (0.412)\\ 
$h$: Cubic        &  {\bf 0.520 (0.299)} &  {\bf 0.564 (0.179) }& {\bf 1.477 (0.406)}\\ 
\hline
FLM              & 2.765 (0.458)& 2.855 (0.440) & 2.858 (0.399) \\  
PAFLM         & 5.021 (4.415)& 5.741 (5.383) & 5.084 (4.703)\\ 
\hline
\multicolumn{4}{||c||}{\bf Nonparametric} \\
\hline
NP-$\ltwo$ & \multicolumn{3}{c||}{1.652 (0.275) } \\
NP-shape  & \multicolumn{3}{c||}{1.960 (0.368) } \\
\hline
\hline
\end{tabular}
\caption{The average and the standard deviation (in parentheses) of RMSEs for three model-based methods on simulated test data. The true values are Fourier2 basis and a quadratic $h$.}
\label{table:simulated}
\end{table}

\subsubsection{Simulated Data 2}
In the second experiment, we again simulate $n = 100$ observations using 
the model stated in Eqn.~\ref{eqn:full-model}, but this time we use a B-spline basis with $20$ elements and 
random coefficients to form the predictor functions.
As earlier, we simulate the coefficient function $\beta$ using the same basis and a fixed coefficient vector.
Then we plug these quantities in the model, use a quadratic polynomial function $h$, and add independent observation noise, $\epsilon_i \sim N(0, 0.01^2)$, to obtain the responses $\{y_i\}$.
Skipping further details, we focus directly on prediction performance 
(using the same B-spline basis with $20$ elements). 

\paragraph{\bf Prediction Performance} The prediction results are shown in Table \ref{table:simulated2}. Despite 
increased complexity of predictors, resulting from a larger basis set, EFRM still performs
better relative to the 
competing methods.

\begin{table}[!htb]
\centering
\begin{tabular}{||c|c|c||}
\hline
\hline
                                        & Model        & RMSE                   \\ \hline \hline
\multirow{5}{*}{\textbf{Parametric}}    & $h$: linear    & 5.984 (2.670)         \\ \cline{2-3} 
                                        & $h$: quadratic & 4.548 (1.703) \\ \cline{2-3} 
                                        & $h$: cubic     & {\bf 4.379 (1.876)}        \\ \cline{2-3} 
                                        & FLM          & 7.698 (1.746)     \\ \cline{2-3} 
                                        & PAFLM        &  36.540 (9.932)         \\ \hline
\multirow{2}{*}{\textbf{Nonparametric}} & NP-$\ltwo$   & 8.969 (1.691)          \\ \cline{2-3} 
                                        & NP-shape   & 10.030 (1.424)          \\ \hline \hline
\end{tabular}
\caption{The average and the standard deviation (in parentheses) of the five RMSE's for three model-based methods on simulated test data}
\label{table:simulated2}
\end{table}

A part of the success of EFRM can be attributed to the fact that the data 
was indeed simulated from that model itself. Therefore, it is natural that this model does 
better than others. However, these experiments also point to the robustness of the response 
variable to random phase variability in the functional predictors. Technically, the response is invariant to this
phase variability. Additionally, the model benefits
from optimization over $\Gamma$ alongside the estimation of $\beta$ and $h$. In this way, 
the model chooses phases in a way that helps maximize prediction performance.

\subsection{Application to Real Data}

Next, we apply EFRM to three real data examples. There are several important application 
areas where functional variables form predictors. Examples 
include biology, human anatomy, biochemistry, finance, epidemiology, and so on. We take three 
representative examples from human biometrics, chemistry and stock market. The goal in each case is to use
shapes of functional predictors in predicting scalar response variables.

\paragraph{\bf Description of the Data}

\begin{enumerate}

\item {\bf Gait in Parkinson's Disease Data}:  First, we use Gait data collected for diagnosing Parkinson's disease, 
taken from the well-known {\it Physionet} \cite{Physiobank} database. The database contains {\it Vertical Ground Reaction Force} (VGRF) records of subjects as they walk at their usual, self-selected pace for approximately two minutes on level ground. A total of 
eight sensors are placed underneath each foot for 
measuring forces (in Newtons) as functions of time. The outputs of these 
$16$ sensors (left: $8$ and right: $8$) 
are digitized and recorded at $100$ samples per second. 
From the original data, we extract very short segments (the first $1-100$ time points from total
$12119$) for simplicity and efficiency of computation.
Based on demographic information, each patient has 
his/her own {\it Timed Up And Go} (TUAG) test which is a simple test used to assess a 
patient's mobility and requires both static and dynamic balance. 
We consider VGRF records as predictors and TUAG as scalar responses, with each subject forming an independent
observation. 

There are three groups of patients in Gait in Parkinson's disease data. We focus on two groups named 
{\it ``Ga''} and {\it ``Si''} \cite{frenkel2005treadmill, frenkel2005effect, yogev2005dual} in the dataset to ensure 
the same demographic information among the participants. This results in a total of $61$ functions or curves for the analysis. 
 
 \begin{figure}[!htb]
    \centering
            \begin{subfigure}[b]{0.45\textwidth}
        \includegraphics[width=\textwidth]{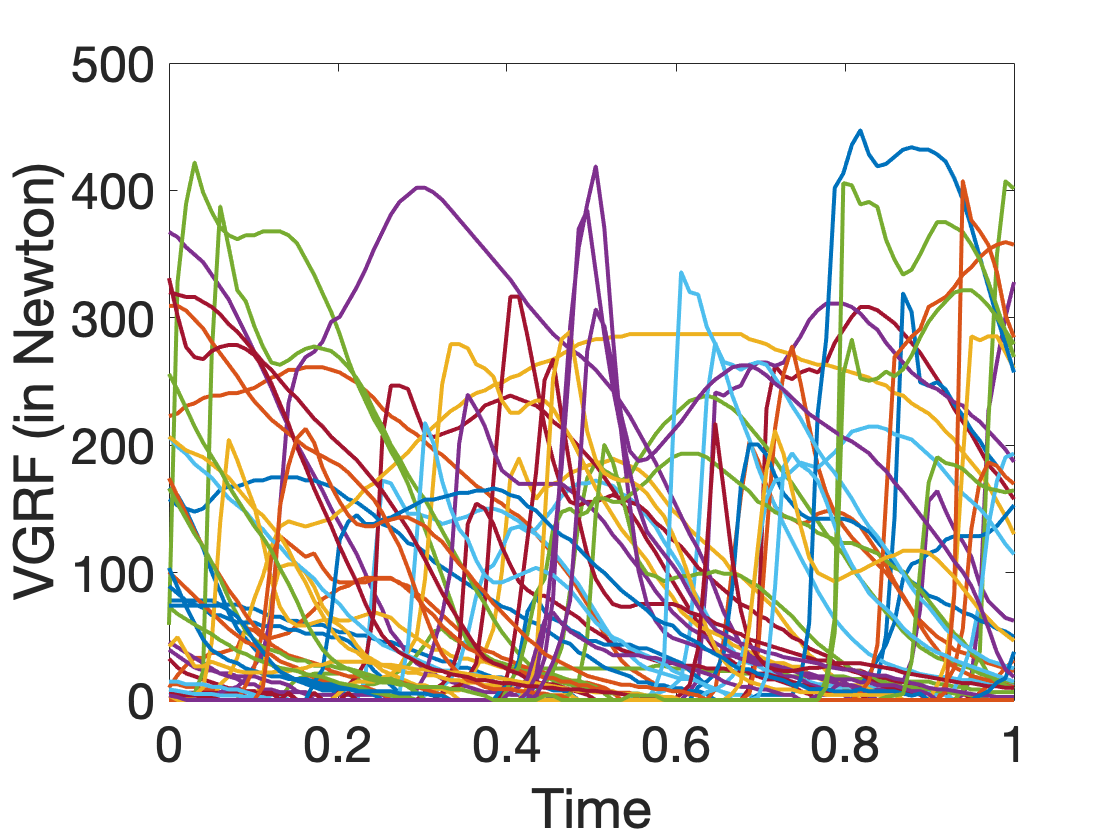} 
       \caption{VGRF, $\{f_i\}$}
        \label{fig:gp}
    \end{subfigure}
    ~ %add desired spacing between images, e. g. ~, \quad, \qquad, \hfill etc. 
    \begin{subfigure}[b]{0.45\textwidth}
        \includegraphics[width=\textwidth]{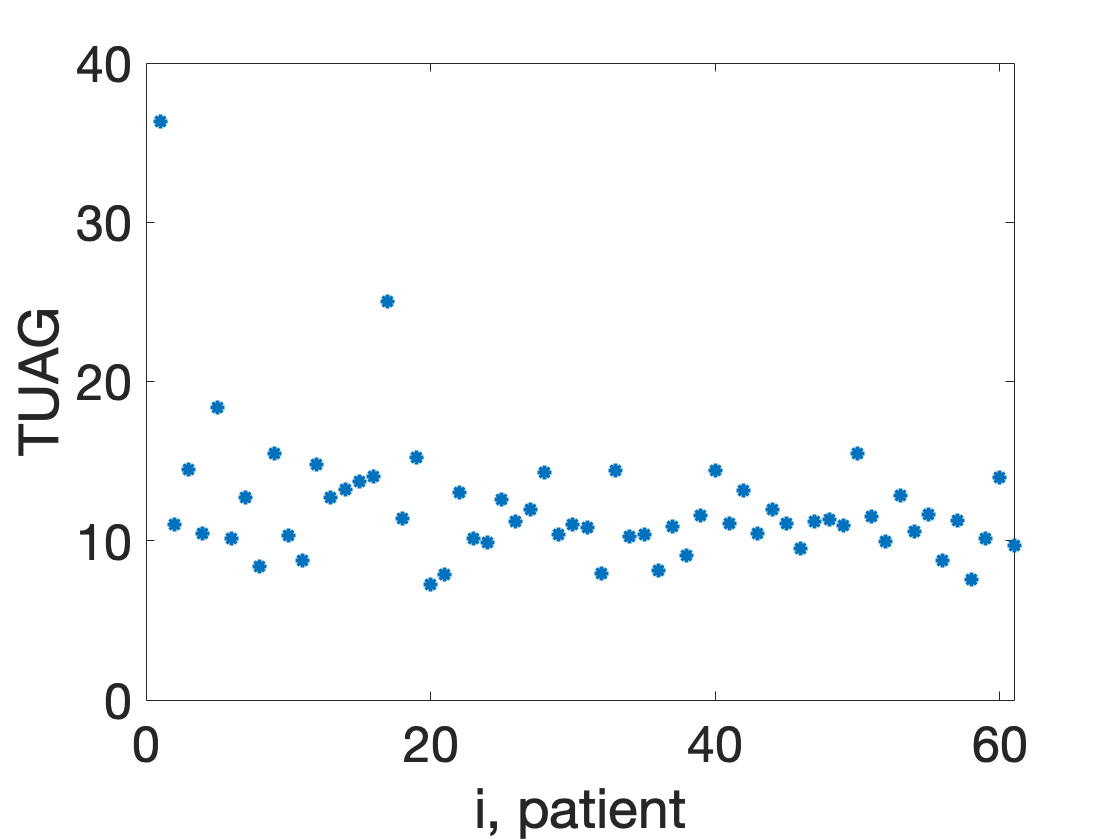} 
        \caption{TUAG, $\{y_i\}$}
        \label{fig:real_y1}
    \end{subfigure}
    \caption{Gait in Parkinson's Disease Data}
    \label{fig:gait}
\end{figure}

Fig. \ref{fig:gait} plots segments of VGRF for each of the 61 patients 
in the left panel and TUAG values in the right panel. 
In the experiments presented later, we randomly select $40$ as training  and the rest $21$ as test.

\item {\bf Metabonomic 1H-NMR Data}: Metabonomic 1H-NMR (Nuclear Magnetic Resonance) data \cite{winning2009exploratory} originates from 1H NMR analysts of urine from thirty-two rats, fed a diet containing an onion by-product. The aim is to evaluate the {\it in vivo} metabolome following 
the intake of onion by-products. The data set contains 31 NMR spectra in the region between $(0 - 3000)$ ppm as predictors and some reference chemical values as responses.

 \begin{figure}[!htb]
    \centering
 %       \captionsetup{font=normalsize,labelfont={bf,sf}}
%   \captionsetup[sub]{font=normalsize,labelfont={bf,sf}}
    \begin{subfigure}[b]{0.45\textwidth}
         \includegraphics[width=\textwidth]{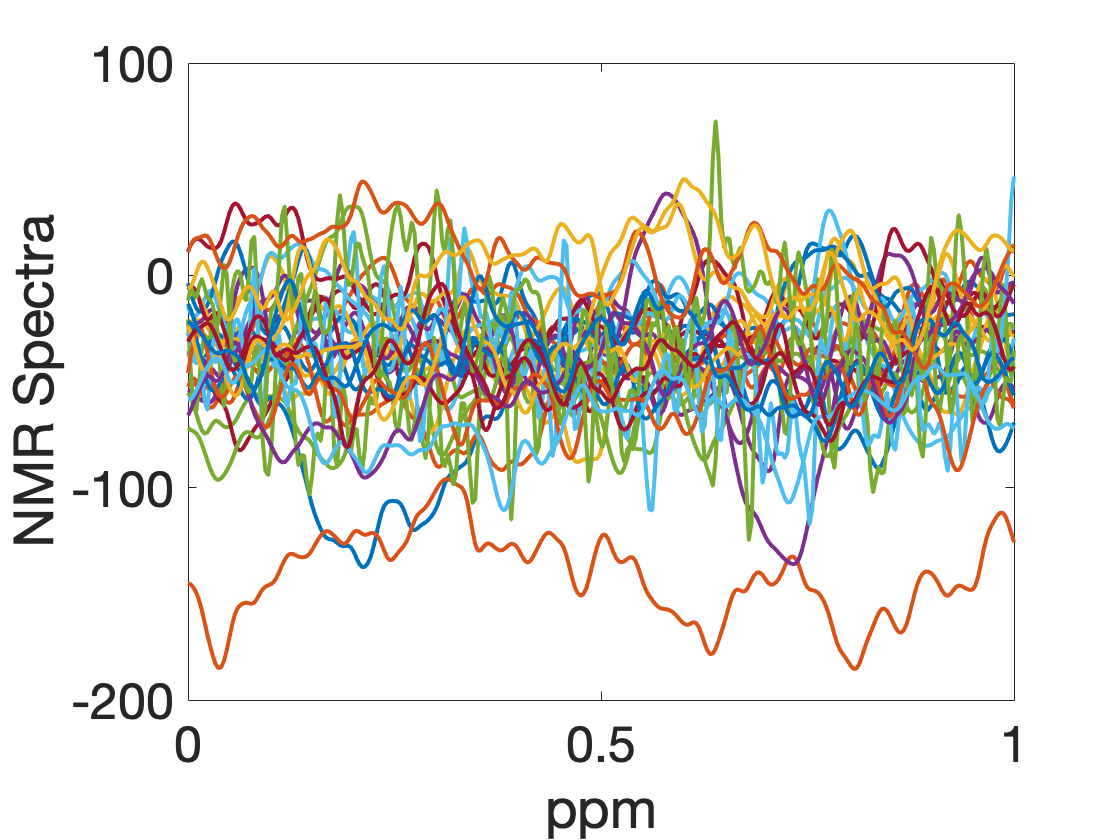} 
       \caption{31 NMR spectra, $\{f_i\}$}
        \label{fig:real_x2}
    \end{subfigure}
    ~ %add desired spacing between images, e. g. ~, \quad, \qquad, \hfill etc. 
    \begin{subfigure}[b]{0.45\textwidth}
        \includegraphics[width=\textwidth]{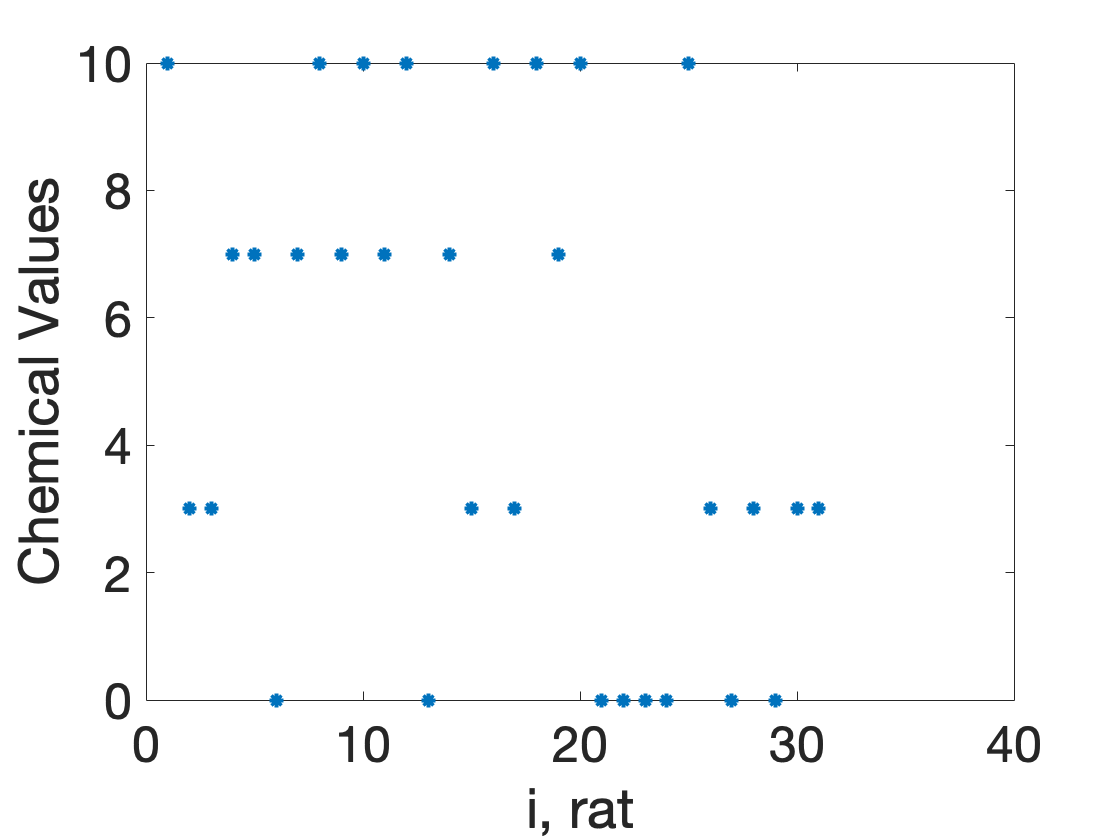} 
        \caption{chemical values, $\{y_i\}$}
        \label{fig:real_y2}
    \end{subfigure}
    \caption{Metabonomic 1H-NMR Data}
    \label{fig:onion}
\end{figure}

Since we have $31$ total observations, we randomly select $21$ curves as the training set and 
rest $10$ curves as the test set. Similar to the Gait in Parkinson's disease data, we extract the first 
$300$ time points from $29001$ time points for efficient 
 computation and statistical analysis. Fig. \ref{fig:onion} displays the plots of NMR spectra of $31$ rats 
 (left panel) and the chemical values which are considered as response variable (right panel). 

\item {\bf Historical Stock Data}: {\it QuantQuote} posts 
large amounts of free historical stock data on their website for free download. 
There are total of $200$ companies and each company has total $3,926$ stock entries
during the interval 1/2/1998 to 8/9/2013. For each company, we collected stock prices from 3/20/2012 to 8/9/2012 
to form functional predictors. Thus, there are
$100$ daily time points over the selected interval forming predictor functions. 
We take the stock prices on 8/9/2013, which is exactly one year after the end of predictor interval, 
as the scalar response variable. Our goal is to predict one-year future stock price for 
each company based on historical stock prices. 

\begin{figure}[!htb]
    \centering
    \begin{subfigure}[b]{0.45\textwidth}
        \includegraphics[width=\textwidth]{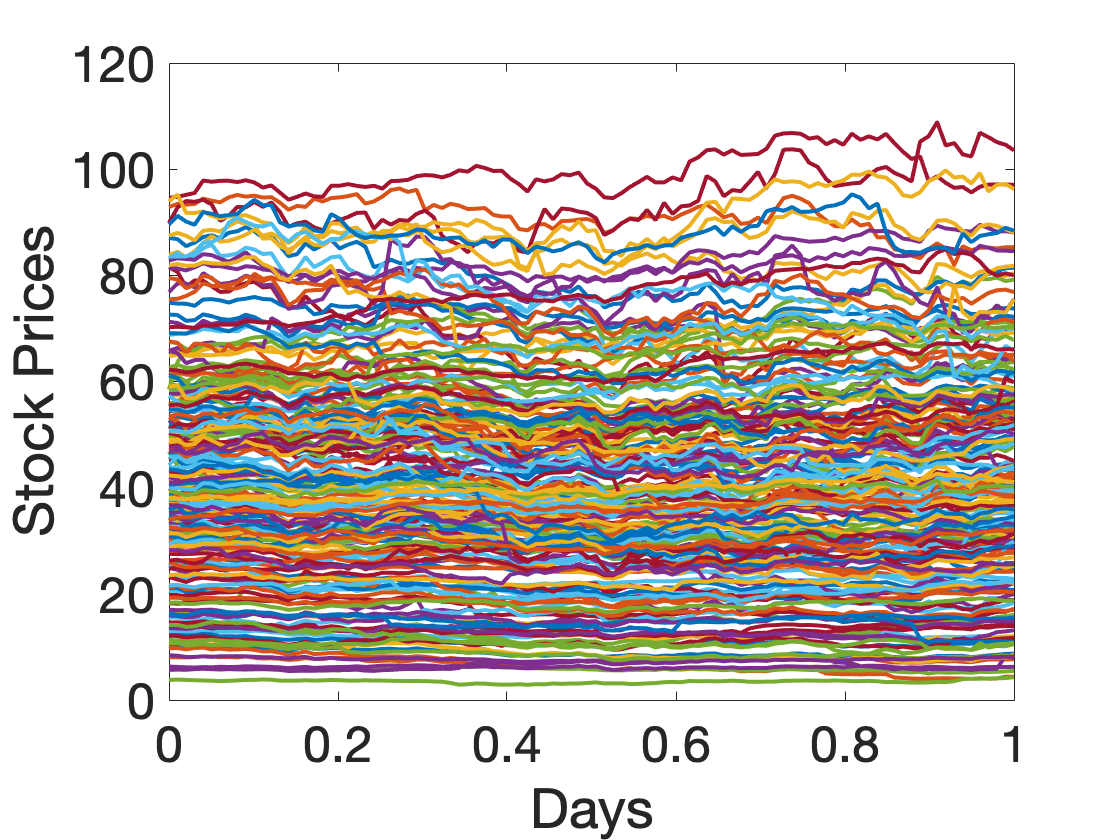} 
       \caption{Past Stock Prices, $\{f_i\}$}
        \label{fig:real_x}
    \end{subfigure}
    ~ %add desired spacing between images, e. g. ~, \quad, \qquad, \hfill etc. 
    \begin{subfigure}[b]{0.45\textwidth}
        \includegraphics[width=\textwidth]{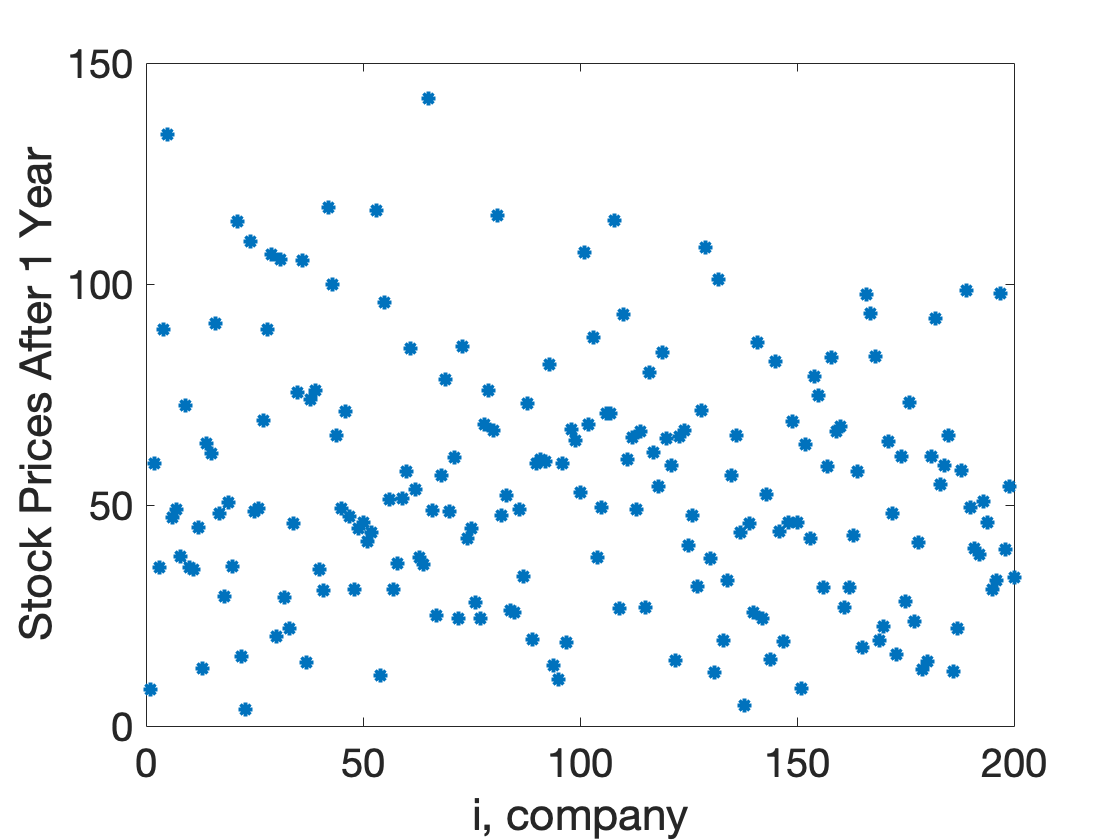} 
        \caption{One-year future value $\{y_i\}$}
        \label{fig:real_y}
    \end{subfigure}
    \caption{Stock Data}
    \label{fig:stock}
\end{figure}

Fig. \ref{fig:stock} shows an example of this stock data. The $200$ functional predictors
are shown in Fig. \ref{fig:real_x} and scalar response variables are shown in Fig. \ref{fig:real_y}. 
We use first $140$ curves to fit the model and remaining $60$ curves as test.

\end{enumerate}

\paragraph{\bf Analysis of Real Data}

For representing the coefficient function $\beta$, 
we use a B-spline basis with $20$ elements and estimate parameters using Algorithm \ref{algo:final}.

\begin{figure}[!]
    \centering
    \begin{subfigure}[b]{.23\textwidth}
         \includegraphics[width=\textwidth]{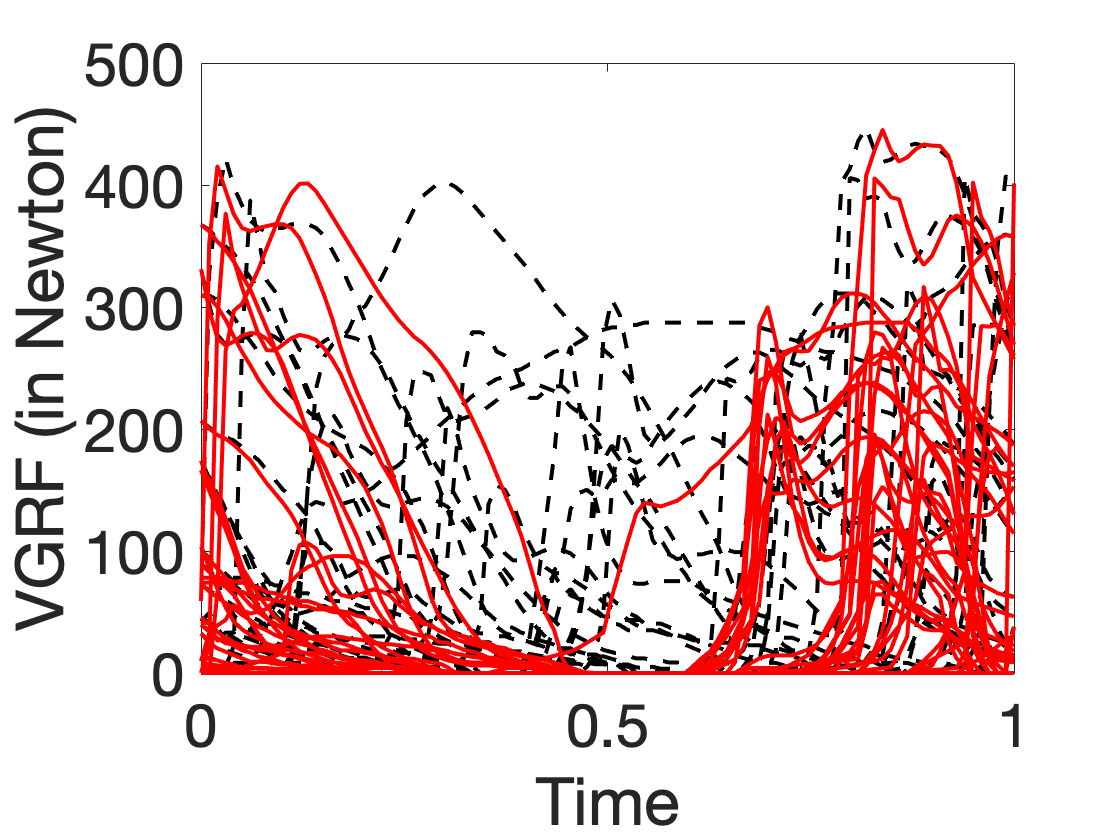} 
        \includegraphics[width=\textwidth]{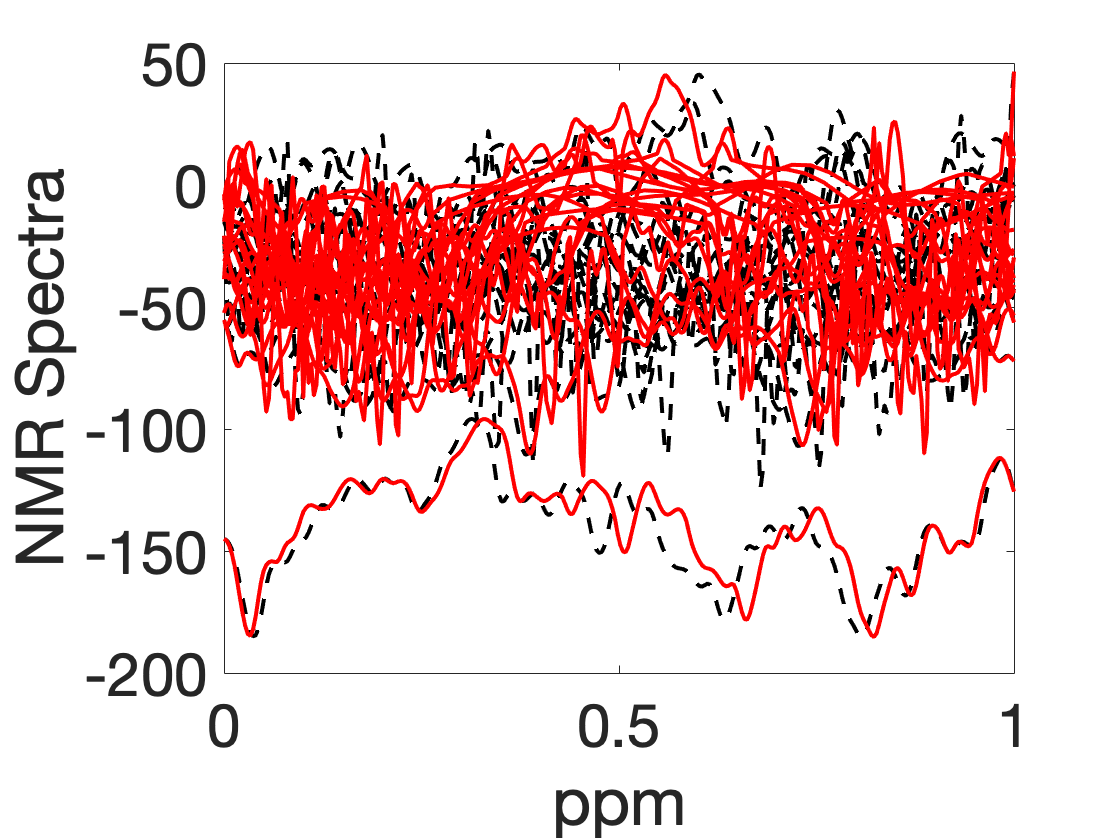} 
        \includegraphics[width=\textwidth]{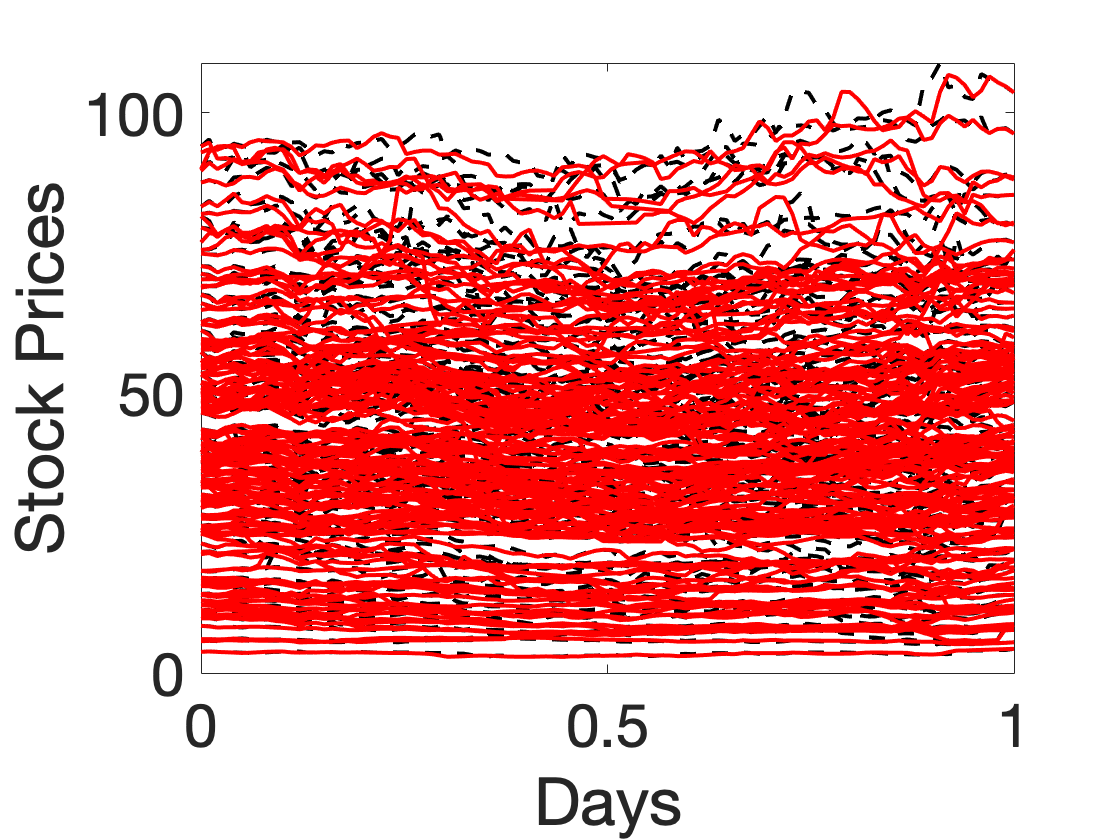} 
       \caption{Training}
        \label{fig:15}
    \end{subfigure}
    ~ %add desired spacing between images, e. g. ~, \quad, \qquad, \hfill etc. 
    \begin{subfigure}[b]{.23\textwidth}
       \includegraphics[width=\textwidth]{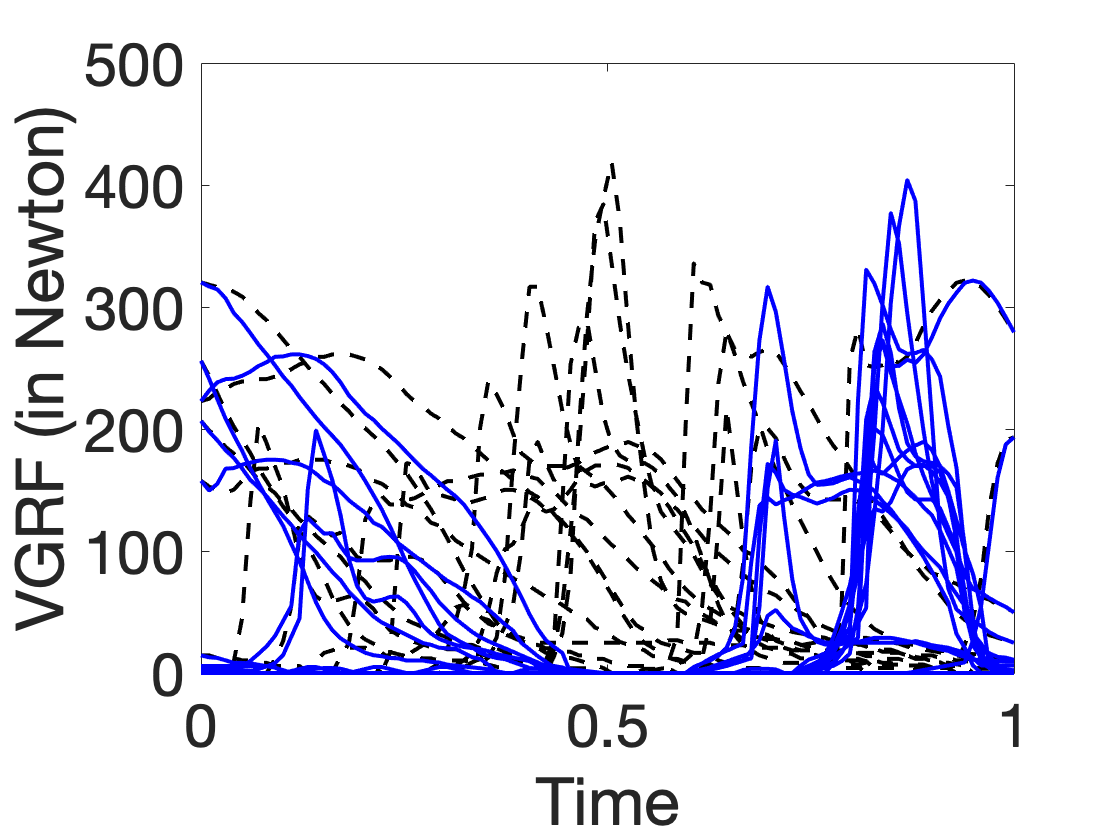} 
        \includegraphics[width=\textwidth]{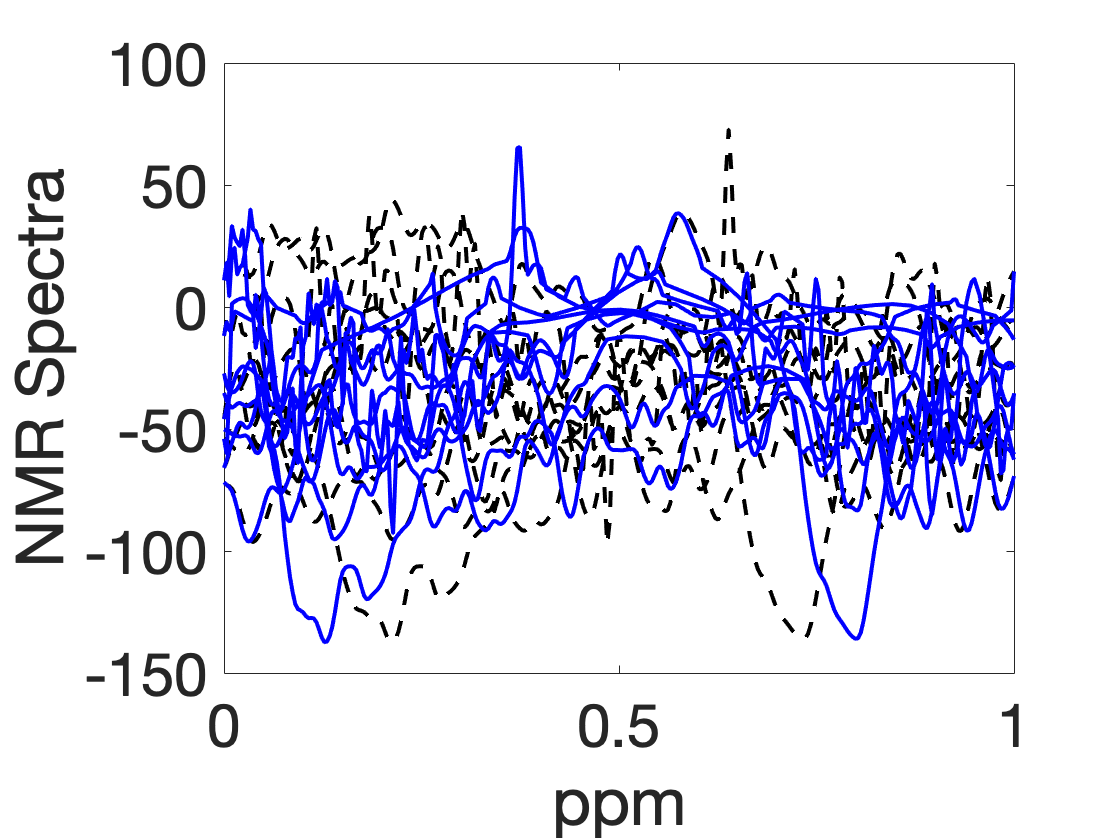} 
        \includegraphics[width=\textwidth]{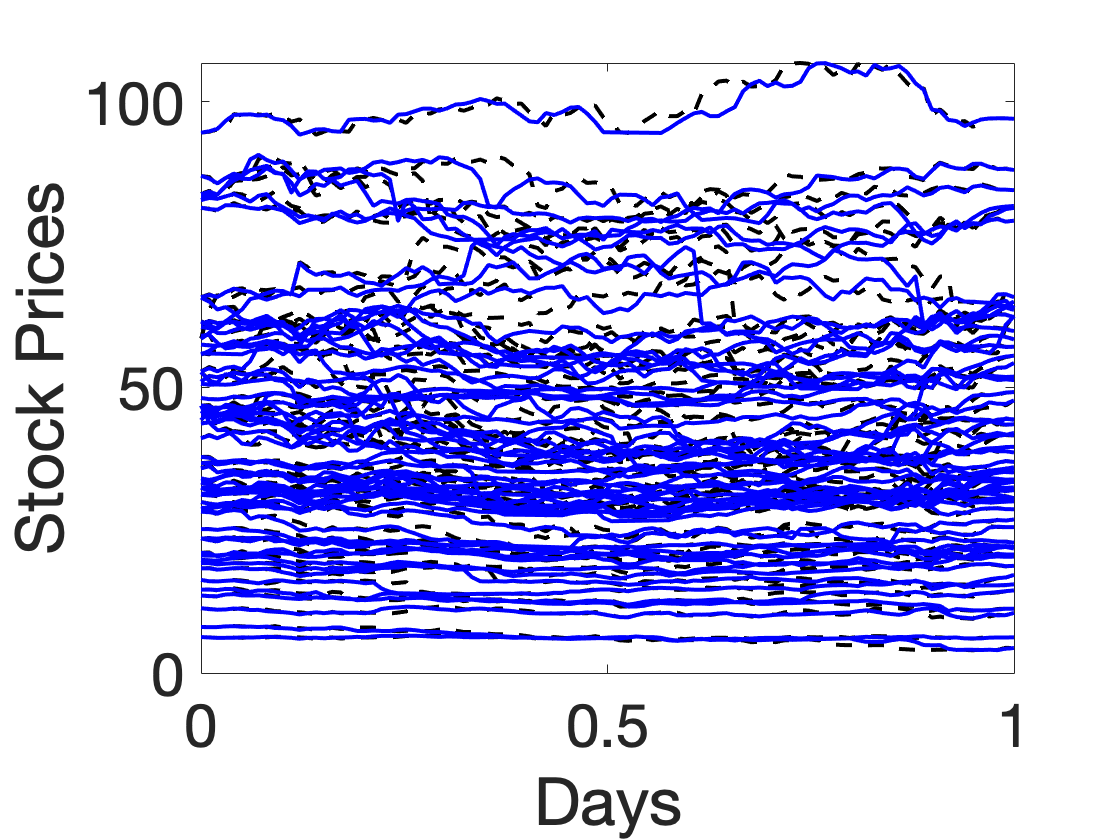} 
        \caption{Test}
        \label{fig:16}
    \end{subfigure}
        \begin{subfigure}[b]{.23\textwidth}
        \includegraphics[width=\textwidth]{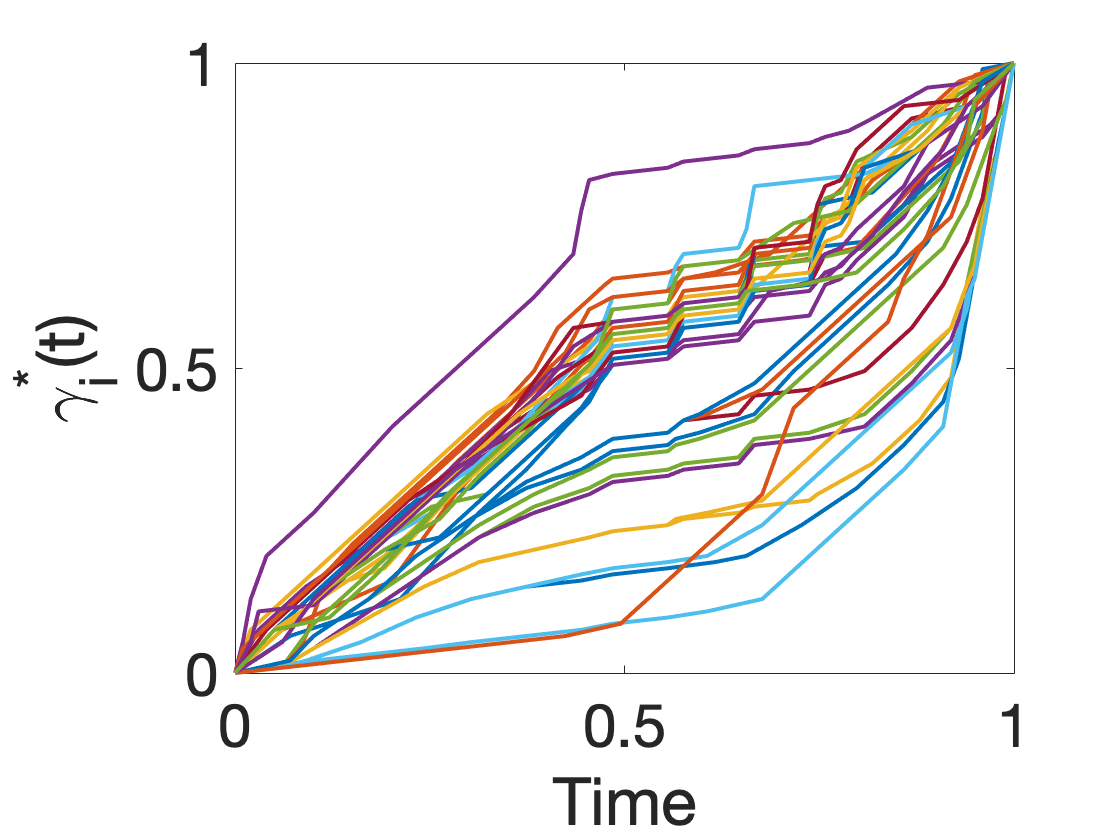} 
        \includegraphics[width=\textwidth]{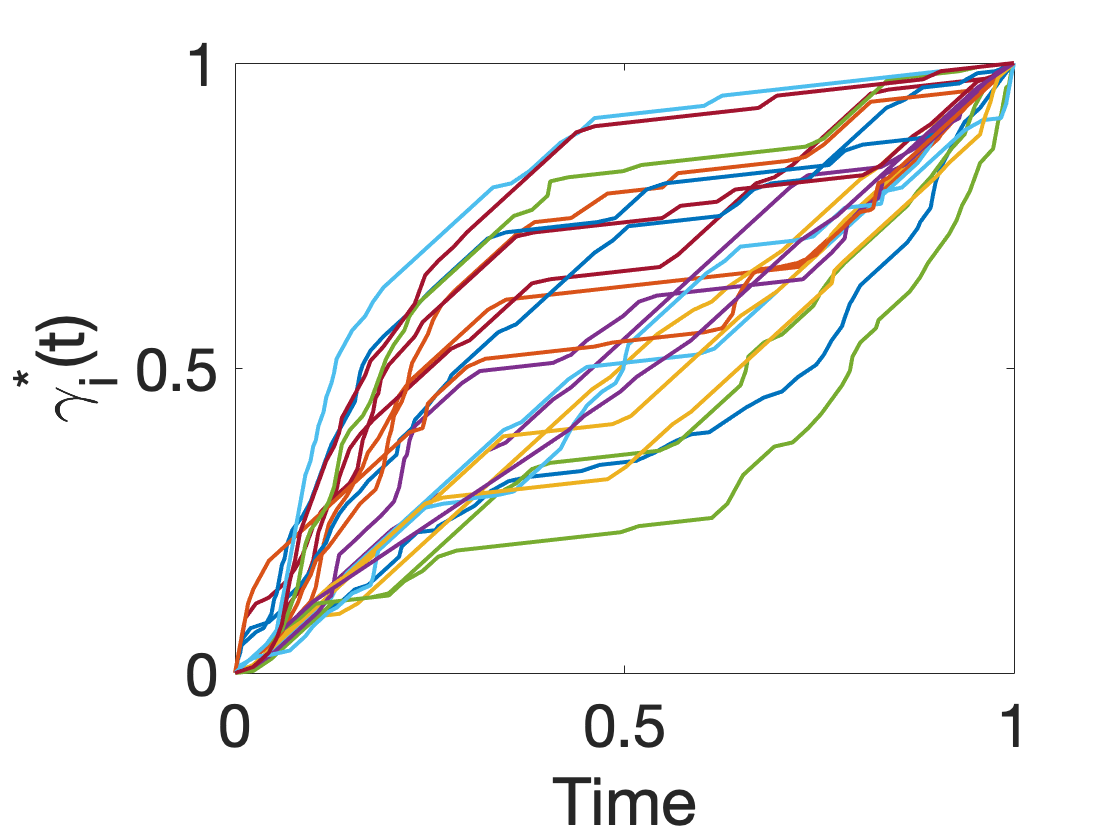} 
        \includegraphics[width=\textwidth]{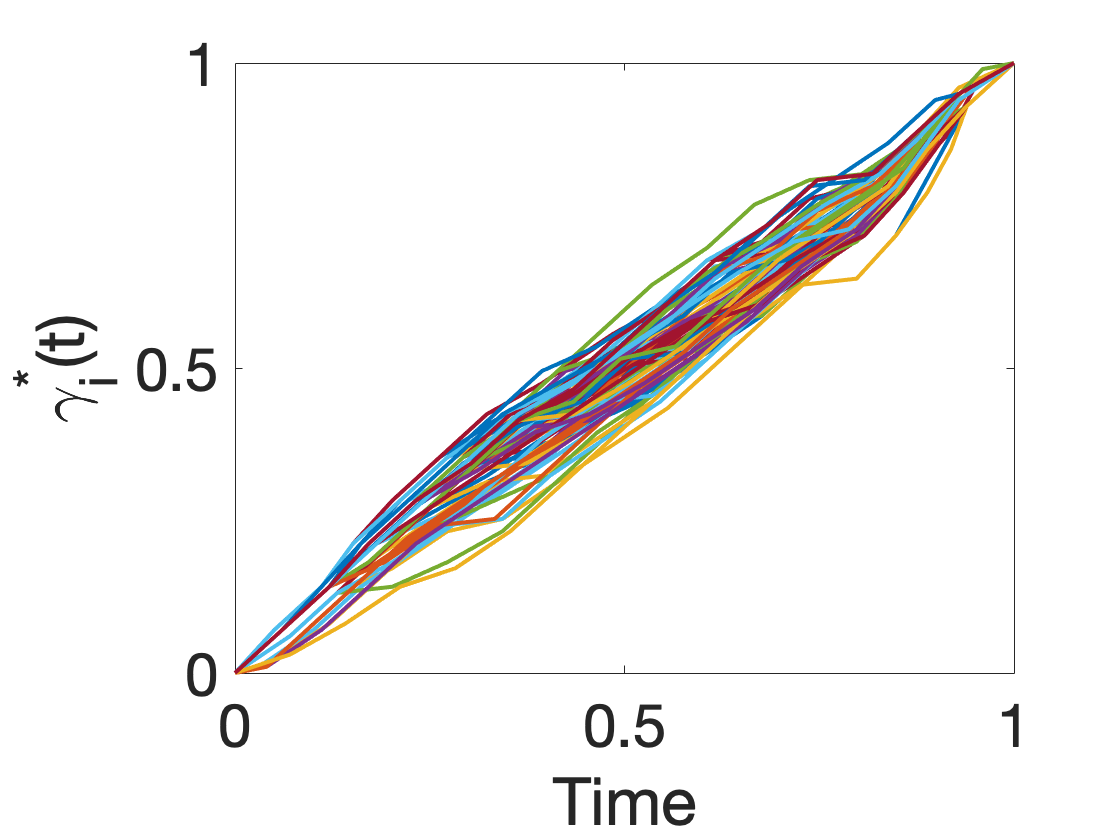} 
        \caption{Training}
        \label{fig:17}
    \end{subfigure}
            \begin{subfigure}[b]{.23\textwidth}
        \includegraphics[width=\textwidth]{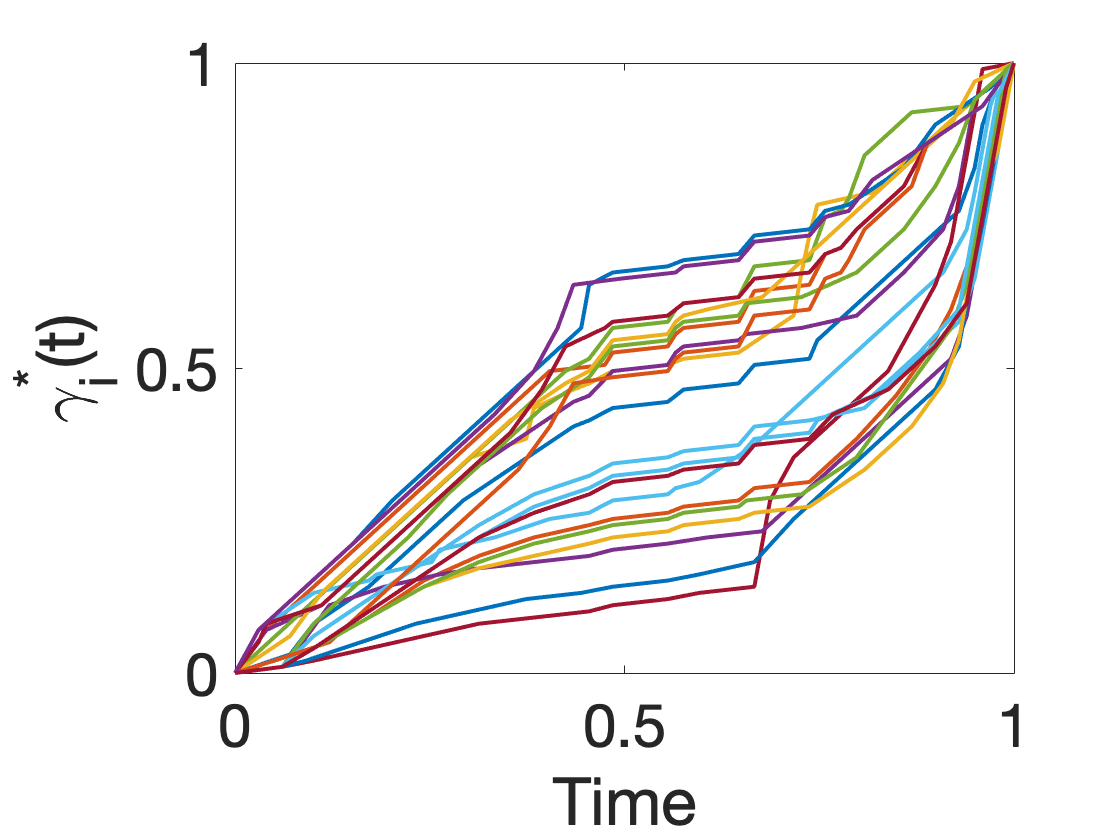} 
         \includegraphics[width=\textwidth]{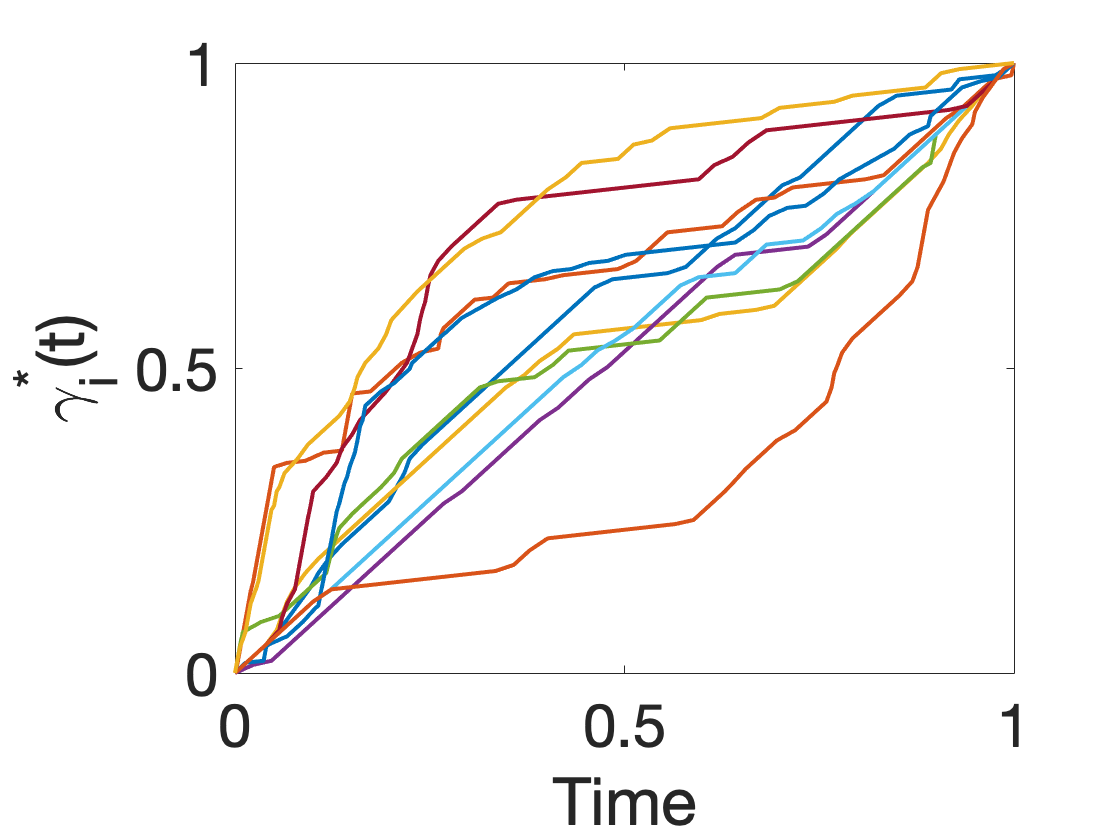} 
        \includegraphics[width=\textwidth]{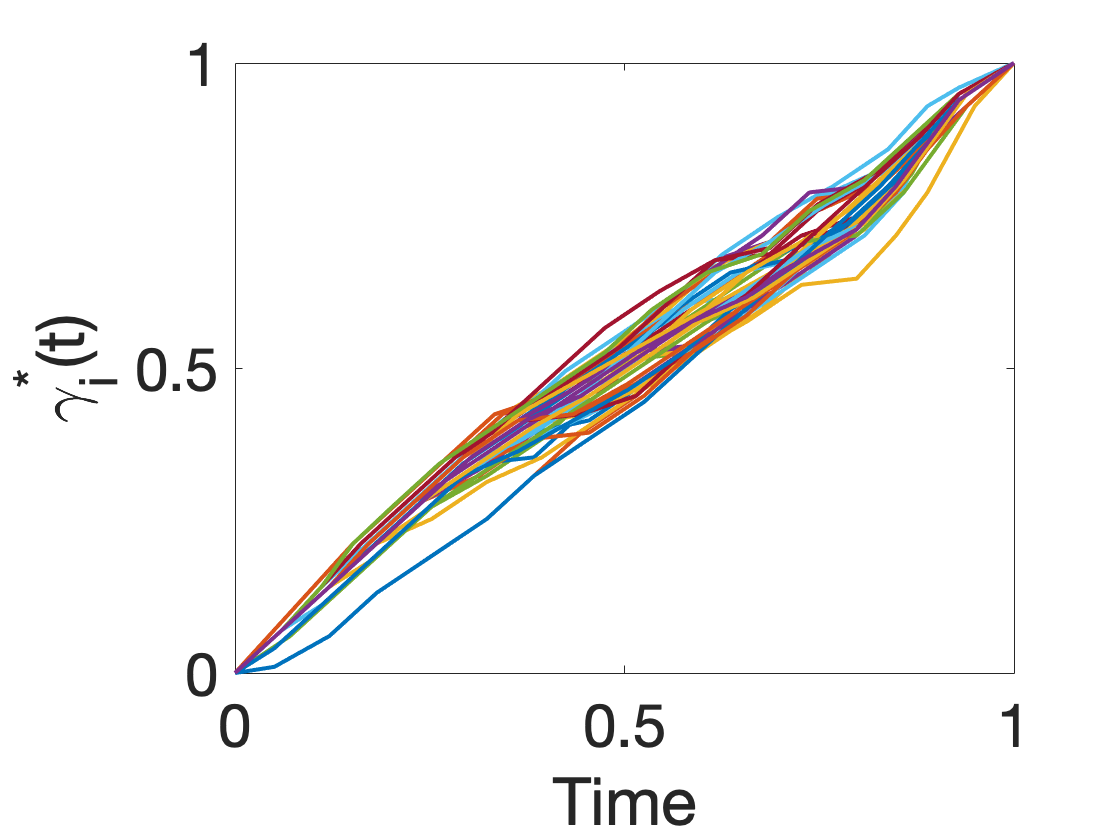} 
        \caption{Test}
        \label{fig:18}
    \end{subfigure}
        \caption{$\{f_i\}$ vs. Warped $\{ f_i\}$ and $\{\gamma^*_i \}$}
    \label{fig:real_estimation}
\end{figure}

Fig. \ref{fig:real_estimation} shows
``aligned" functional predictors during 
training and testing. Each row corresponds to a real data set
-- gait in Parkinson's disease (first row), metabonomic 1H-NMR (second row), and historical stock market (third row).
The original functions are drawn in black dashed curves and the warped functions are overlaid
using the red/blue solid colors. Fig. \ref{fig:15} and \ref{fig:16} show the curves for the training data and the test data, respectively. 
The corresponding optimal warpings for training and test are shown in Fig. \ref{fig:17} and \ref{fig:18}, respectively. 
We remind the readers that the predictors have been warped using the norm-preserving action
during the optimization step. They appear more aligned than before but 
are not as aligned as one would get from a pure alignment procedure. 
This alignment results in an 
increased ability of the model to predict the response variable. Thus, this warping is more 
to help regress the responses $y_i$s to the predictors $f_i$s, rather than to align peaks and valleys in 
$f_i$s.

\paragraph{\bf Prediction Results}

Table \ref{table:real_result_total} presents prediction RMSE for different models 
studied in this experiment. It shows that EFRM model outperforms other models on 
all three datasets.  In the case of 1H-NMR data, EFRM using a cubic index function does the best, 
while in other cases lower order polynomials perform better. 
This could be because the response variable in NMR example is categorical with four values
and one needs a cubic polynomial to fit these response levels.
Predictions from the kernel regression model are close second to EFRM. 
%The exception may be because observed functions have different heights and starting points. Functional predictors in each training data and test data have different shapes (different heights and starting points) so nonparametric method cannot handle this problem.
\begin{table}[htb!]
\centering
\begin{tabular}{||c||c|c|c||}
\hline
\hline
 Model  & Gait  & 1H-NMR & Stock\\
 \hline
 \hline
 $h$ : Linear & {2.741}          & 4.849       & {\bf 9.007}\\
 $h$ : Quadratic & {\bf 2.466} & 4.106       & 9.130\\
 $h$ : Cubic & 2.594             & {\bf 4.025} & 9.227\\
 \hline
 FLM & 7.483 & 213.490 & 10.405\\
 PAFLM & 19.158 & 202.941 & 11.086\\
 NP-$\ltwo$& 6.559 & 4.251 & 9.795 \\
 NP-shape & 2.625 & 4.251 & 9.540\\
 \hline
 \hline
\end{tabular}
\caption{Prediction RMSE for predicted response variable under each model.}
\label{table:real_result_total}
\end{table}

\section{Concluding Remarks}

The development of functional regression models that can handle 
phase variability in functional predictors is a challenging problem in FDA. 
We have proposed a new elastic approach that uses the shapes of functions, rather than the full functions, 
as predictors in regression models. 
The notion of shape is based on a norm-preserving warping 
of the predictors and handles the nuisance phase variability by 
optimizing the $\ltwo$ inner product over the warping group inside the model. 
We compare the prediction RMSE of the model with several existing methods, 
to demonstrate effectiveness of this technique in both simulated data and real data. 

%We emphasize that while phase is nuisance in some applications, it is not always
%the case. One should not expect shapes of predictors to be predominant in 
%all situations. Phase components may also carry important information about the responses and one can not always ignore them. However, in some cases, as illustrated through the simulated data and real data examples presented in this paper, shape can be the primary predictor and one wants 
%models that can exploit that knowledge.

As discussed in Section 1.3, there is another model that can potentially eliminate the effects of 
phase variability in the predictor functional data. This model involves SRVFs $\{q_i\}$ of the 
predictors and uses the term $\sup_{\gamma_i} \inner{(q_i \circ \gamma_i)\sqrt{\dot{\gamma}_i}}{\beta}$
as the argument of the index function $h$. However, we have not pursued this model because, 
despite theoretical advantages, the practical performances of this model are sometimes low. As an example, 
we study the prediction problem using the same stock market data as in item 3 of Section 3.2. 
The prediction RMSE for this model is listed in Table \ref{table:prediction_srvf}, and is found to be 
worse than the results shown in Table \ref{table:real_result_total}. We conjecture that it is because 
the noise in predictor data gets enhanced when computing SRVFs (due to the presence of a time 
derivative in SRVF expression). Thus, we prefer the second option mentioned in Section 1.3 for 
EFRM. 

\begin{table}[!htb]
\centering
\begin{tabular}{||c|c|c|c||}
\hline \hline
Model & $h$: Linear & $h$: Quadratic   & $h$: Cubic     \\ \hline \hline
RMSE  & 18.032 & 18.139 & 18.110   \\ \hline \hline
\end{tabular}
\caption{RMSEs of using SRVF representation and value-preserving warping.}
\label{table:prediction_srvf}
\end{table}

\section*{Acknowledgement}

This research was supported in part by the NSF grants NSF-1621787 and NSF-1617397 to AS. 
This paper describes objective technical results and analysis. Any subjective views
or opinions that might be expressed in the paper do not necessarily represent the views
of the U.S. Department of Energy or the United States Government. Supported by
the Laboratory Directed Research and Development program at Sandia National Laboratories, a multi-mission laboratory managed and operated by National Technology
and Engineering Solutions of Sandia, LLC, a wholly owned subsidiary of Honeywell
International, Inc., for the U.S. Department of Energy's National Nuclear Security Administration under contract DE-NA0003525.

\section*{Availability}
MATLAB programs are available on Github: \href{https://github.com/fdastat/elastic-regression}{https://github.com/fdastat/elastic-regression}

\section*{References}
\bibliography{MKAbib}
\end{document}